\documentstyle[psfig]{mn}
\input mnras.add
\makeatletter
\newcounter{cureqno}
\newenvironment{mathletters}{\refstepcounter{equation}%
    \setcounter{cureqno}{\value{equation}}%
    \let\@curtheeqn\theequation%
    \edef\cur@eqn{\theequation}%
    \def\theequation{\cur@eqn\alph{equation}}%
    \setcounter{equation}{0}}%
    {\let\theequation\@curtheeqn%
    \setcounter{equation}{\value{cureqno}}}
\def\figsize{\ifSFB@referee0.5\hsize\else\hsize\fi}
\makeatother
\def\eq#1{\begin{equation} #1 \end{equation}}
\def\eqarray#1{\begin{eqnarray} #1 \end{eqnarray}}

\def\non    {\nonumber \\}
\def\DS     {\displaystyle}
\def\Ref    {\item}
\def\E#1{\hbox{$10^{#1}$}}
\def\sub#1{_{\rm #1}}
\def\case#1/#2{\hbox{$\frac{#1}{#2}$}}
\def\about  {\hbox{$\sim$}}
\def\x      {\hbox{$\times$}}
\def\PsiO   {\hbox{$\Psi_0$}}
\def\Q      {\hbox{$Q\sub{pr,\lambda}$}}
\def\Qi     {\hbox{$Q\sub{pr,i\lambda}$}}
\def\QP     {\hbox{$Q\sub P$}}

\def\QV     {\hbox{$Q\sub V$}}
\def\Qast   {\hbox{$Q_\ast$}}
\def\tV     {\hbox{$\tau\sub V$}}
\def\tF     {\hbox{$\tau\sub F$}}
\def\vf     {\hbox{$v_\infty$}}
\def\vff    {\hbox{$v_\infty^2$}}
\def\vfff   {\hbox{$v_\infty^3$}}
\def\wf     {\hbox{$w_\infty$}}
\def\wff    {\hbox{$w_\infty^2$}}
\def\wfff   {\hbox{$w_\infty^3$}}
\def\vp     {\hbox{$v\sub p$}}
\def\vpp    {\hbox{$v^2\sub p$}}
\def\vD     {\hbox{$v\sub D$}}
\def\vDD    {\hbox{$v^2\sub D$}}
\def\vm     {\hbox{$v\sub m$}}
\def\vg     {\hbox{$v\sub g$}}
\def\vgg    {\hbox{$v^2\sub g$}}
\def\vT     {\hbox{$v\sub T$}}
\def\vTT    {\hbox{$v^2\sub T$}}
\def\vrel   {\hbox{$v\sub{rel}$}}
\def\sg     {\hbox{$\sigma\sub g$}}
\def\ss     {\hbox{$\sigma_{22}$}}
\def\ssq    {\hbox{$\sigma_{22}^2$}}
\def\kms    {\hbox{km s$^{-1}$}}
\def\vd     {\hbox{$v\sub d$}}
\def\nd     {\hbox{$n\sub d$}}
\def\nH     {\hbox{$n\sub H$}}
\def\mp     {\hbox{$m\sub p$}}
\def\Mdot   {\hbox{$\dot M$}}
\def\Msix   {\hbox{$\dot M_{-6}$}}
\def\Mc     {\hbox{$\dot M\sub c$}}
\def\Lo     {\hbox{L$_\odot$}}
\def\Mo     {\hbox{M$_\odot$}}
\def\mic    {\hbox{$\umu$m}}

\def\Pmax   {\hbox{$P\sub{max}$}}
\def\Tc     {\hbox{$T\sub{c}$}}
\def\Tct    {\hbox{$T\sub{c3}$}}
\def\Td     {\hbox{$T\sub{d}$}}
\def\Tk     {\hbox{$T\sub{k}$}}
\def\Tkt    {\hbox{$T\sub{k3}$}}
\def\gm     {\hbox{$g\sub{max}$}}
\def\rdg    {\hbox{$r\sub{dg}$}}
\def\am     {\hbox{$a\sub{min}$}}
\def\dmax   {\hbox{$\delta\sub{max}$}}
\def\xi     {\hbox{$x_{1i}$}}
\def\fbar   {\hbox{$\bar \phi$}}

\pagerange{\pageref{firstpage}--\pageref{lastpage}} \pubyear{2001}

\title[Dusty Winds I]
              {Dusty winds: I. Self-similar solutions}

\author[Moshe Elitzur \& \v{Z}eljko\ Ivezi\'c]
       {Moshe Elitzur$^1$ and \v{Z}eljko Ivezi\'c$^2$\\
   $^1$Department of Physics and Astronomy,
       University of Kentucky,
       Lexington, KY 40506-0055,
       USA;
       moshe@pa.uky.edu\\
   $^2$Department of Astrophysical Sciences,
       Princeton University,
       Princeton, NJ 08544-1001,
       USA;
       ivezic@astro.princeton.edu\\}

\date{Accepted May 15, 2001. Received May 6, 2001; in original form September 12, 2000}
\begin{document}

\maketitle

\label{firstpage}

\begin{abstract}
We address the dusty wind problem, from the point where dust formation has been
completed and outward. Given grain properties, both radiative transfer and
hydrodynamics components of the problem are fully defined by four additional
input parameters. The wind radiative emission and the shape of its velocity
profile are both independent of the actual magnitude of the velocity and are
determined by just three dimensionless free parameters. Of the three, only one
is always significant---for most of phase space the solution is described by a
set of similarity functions of a single independent variable, which can be
chosen as the overall optical depth at visual \tV. The self-similarity implies
general scaling relations among mass loss rate (\Mdot), luminosity ($L$) and
terminal velocity (\vf). Systems with different \Mdot, $L$ and \vf\ but the
same combination $\Mdot/L^{3/4}$ necessarily have also the same $\Mdot\vf/L$.
For optically thin winds we find the exact analytic solution, including the
effects of radiation pressure, gravitation and (sub- and supersonic) dust
drift.  For optically thick winds we present numerical results that cover the
entire relevant range of optical depths, and summarize all correlations among
the three global parameters in terms of \tV. In all winds, $\Mdot \propto
\vfff(1 + \tV)^{1.5}$ with a proportionality constant that depends only on
grain properties. The optically thin end of this universal correlation, $\Mdot
\propto \vfff$, has been verified in observations; even though the wind is
driven by radiation pressure, the luminosity does not enter because of the
dominant role of dust drift in this regime. The \Mdot--$L$ correlation is
$\Mdot \propto (L\tV)^{3/4}(1 + \tV)^{0.105}$. At a fixed luminosity, \Mdot\ is
{\em not} linearly proportional to \tV, again because of dust drift. The
velocity-luminosity correlation is $\vf \propto (L\tV)^{1/4}(1 +
\tV)^{-0.465}$, explaining the narrow range of outflow velocities displayed by
dusty winds. Eliminating \tV\ produces $\vfff = A\,\Mdot\left(1 +
B\,\Mdot^{4/3}/L\right)^{-1.5}$, where $A$ and $B$ are coefficients that
contain the only dependence of this universal correlation on chemical
composition.  At a given $L$, the maximal velocity of a dusty wind is
$v\sub{max} \propto L^{1/4}$ attained at $\Mdot \propto L^{3/4}$, with
proportionality coefficients derived from $A$ and $B$.

\end{abstract}


\section{INTRODUCTION}

Stars on the asymptotic giant branch (AGB) make a strong impact on the galactic
environment.  Stellar winds blown during this evolutionary phase are an
important component of mass return into the interstellar medium and may account
for a significant fraction of interstellar dust.  These dusty winds reprocess
the stellar radiation, shifting the spectral shape toward the infrared. There
are good indications that they dominate the IR signature of normal elliptical
galaxies (Knapp, Gunn \& Wynn-Williams, 1992). In addition to its obvious
significance for the theory of stellar evolution, the study of AGB winds has
important implications for the structure and evolution of galaxies.

Because of the reddening, dusty winds are best studied in the infrared. Since
the observed radiation has undergone significant processing in the surrounding
dust shell, interpretation of the observations necessitates detailed radiative
transfer calculations. The complexity of these calculations is compounded by
the fact that the wind driving force is radiation pressure on the grains;
complete calculations require a solution of the coupled hydrodynamics and
radiative transfer problems. Traditionally these calculations involved a large
number of input parameters that fall into two categories.  The first one
involves the dust properties; widely employed quantities include the dust
abundance, grains size distribution, solid density, condensation temperature
and absorption and scattering efficiencies. The second category involves global
properties, including the stellar temperature $T_\ast$, luminosity $L$ =
\E4$L_4$ \Lo, mass $M = M_0$ \Mo\ and the mass-loss rate \Mdot\ = \E{-6}\Msix\
\Mo\ yr$^{-1}$.

The large number of input quantities complicates modeling efforts, making it
unclear what are the truly independent parameters and which properties can
actually be determined from a given set of data. In a previous study we noted
that the dusty wind problem possesses general scaling properties such that, for
a given type of grains, both the dynamics and radiative transfer depend chiefly
on a single parameter --- the overall optical depth (Ivezi\'c \& Elitzur 1995,
hereafter IE95).  In a subsequent study we established in full rigor the
scaling properties of the dust radiative transfer problem under the most
general, arbitrary circumstances (Ivezi\'c \& Elitzur 1997, hereafter IE97).
Here we extend rigorous scaling analysis to the other aspect of the dusty wind
problem, the dynamics.

In scaling analysis the basic equations are reduced to the minimal number of
free parameters that are truly independent of each other. By its nature, such
analysis is driven by the underlying mathematics, and the proper free
parameters are not necessarily convenient for handling the data. Attempting to
make our presentation tractable we have separated it into two parts.  The
present paper discusses all the theoretical and mathematical aspects of the
dusty wind problem. In a companion paper (Ivezi\'c \& Elitzur 2001, paper II
hereafter) written as a stand-alone, the observational implications are
discussed separately on their own. Readers mostly interested in practical
applications may proceed directly to paper II.

\section{UNDERLYING THEORY}
\label{Theory}

\subsection{Problem Overview}
The complete description of a dusty wind should start at its origin, the
stellar atmosphere.  Beginning with a full atmospheric model, it should
incorporate the processes that initiate the outflow and set the value of \Mdot.
These processes are yet to be identified with certainty, the most promising are
stellar pulsation (e.g. Bowen 1989) and radiation pressure on the water
molecules (e.g. Elitzur, Brown \& Johnson 1989). Proper description of these
processes should be followed by that for grain formation and growth, and
subsequent wind dynamics.

An ambitious program attempting to incorporate as many aspects of this
formidable task as possible has been conducted over the past few years,
yielding models in qualitative agreement with observations (see Fleischer,
Winters \& Sedlmayr 1999 and H\"{o}fner 1999, and references therein). However,
the complexity of this undertaking makes it difficult to assess the meaning of
its successes. In spite of continuous progress, detailed understanding of
atmospheric dynamics and grain formation is still far from complete. When
involved models succeed in spite of the many uncertain ingredients they
contain, it is not clear whether these ingredients were properly accounted for
or are simply irrelevant to the final outcome.

Fortunately, the full problem splits naturally to two parts, as recognized long
ago by Goldreich \& Scoville (1976, GS hereafter). Once radiation pressure on
the dust grains exceeds all other forces, the rapid acceleration to supersonic
velocities produces complete decoupling from the earlier phases that contain
all the major uncertainties. Subsequent stages of the outflow are independent
of the details of dust formation---they depend only on the final properties of
the grains, not on how these grains were produced; the supersonic phase would
be exactly the same in two different outflows if they have the same mass-loss
rate and grain properties even if the grains were produced by entirely
different processes. Furthermore, these stages are controlled by processes that
are much less dependent on detailed micro-physics, and are reasonably well
understood. And since most observations probe only the supersonic phase, models
devoted exclusively to this stage should reproduce the observable results
while avoiding the pitfalls and uncertainties of dust formation and the wind
initiation. For these reasons, the GS approach with its focus on the supersonic
phase has been widely used in studies of the dusty wind problem (including
recent ones by Netzer \& Elitzur 1993, NE hereafter, and Habing, Tignon \&
Tielens 1994, HTT hereafter). This is the problem we address here.

\subsubsection{Overall Plan}
We consider a spherical wind in steady state (the steady-state assumption is
adequate as long as the wind structure is not resolved in too fine details; see
IE95). Our starting point is the radius $r_1$ beyond which the properties of
individual dust grains do not change and radiation pressure is the dominant
force on the envelope. When positions in the shell are specified in terms of
the scaled radius $y = r/r_1$, the shell inner boundary is always at $y = 1$
and the actual magnitude of $r_1$ drops out of the problem. The equation of
motion is $\rho dv/dt = \cal F$, where $\cal F$ is the net outward radial force
per unit volume, $v$ is the gas velocity and $\rho = \nH\mp$ is its density
(\nH\ is the number density of hydrogen nuclei and \mp\ is the proton mass). In
steady-state $dt = dr/v$ and the equation becomes
\eq{\label{force - Ek}
  {dv^2\over dy} = 2a_{\cal F}r_1,
}
where $a_{\cal F} = {\cal F/\rho}$ is the acceleration associated with force
$\cal F$. Since $a_{\cal F}r_1$ has dimensions of $v^2$, any force can always
be characterized by the velocity scale it introduces.

Winds of interest are highly supersonic, therefore the gas pressure gradient
can be neglected (see also \S \ref{sec:boundaries}). The expansion is driven by
radiation pressure on the dust grains and is opposed by the gravitational pull
of the star. The dust and gas particles are coupled by the internal drag force.
For each of these three force components we first derive the characteristic
velocity scale and the dimensionless profile associated with its radial
variation. With the resulting expressions we identify all the dimensionless
free parameters and formulate the problem in terms of independent dimensionless
variables, resulting in two coupled equations for the dynamics and radiative
transfer. We proceed to solve the mathematical problem, and afterwards
transform the dimensionless free parameters back into the physical variables
that characterize the system.  This procedure ensures that its outcome contains
all the correlations that exist among the physical parameters of dusty winds.

Our presentation starts with a single type of dust grains, section
\ref{Mixtures} extends the discussion to mixtures of sizes and chemical
compositions. The grain is specified by its size $a$, condensation temperature
\Tc\ and absorption and scattering efficiencies $Q\sub{abs,\lambda}$ and
$Q\sub{sca,\lambda}$. We associate the radius $r_1$ with prompt dust formation
so that $\Td(r_1)$ = \Tc. The mathematical problem does not contain any
reference to $r_1$, its actual magnitude enters only during the final
transformation to physical quantities. We address this issue again in our
summary in section \ref{summary}. Table 1 lists the dust properties used in our
numerical applications. Appendix \ref{app:Glossary} contains a glossary of all
the relevant symbols.

%%%%%%%%%%%%%%%%%%%%%%%%%%%%% Table 1 %%%%%%%%%%%%%%%%%%%%%%%%%%%%%
\begin{table}
\begin{center}
\begin{tabular}{lrr}
\hline
               &  Carbon & Silicate  \\
\hline \hline
   $a$ (\mic)  &   0.1   &   0.1  \\
   \Tc\ (K)    &   800   &   800  \\
   \hline
   \QV         &  2.40   &  1.15  \\
   \Qast       &  .599   &  .114  \\
   \PsiO       &  5.97   &  2.72  \\
   \hline
\end{tabular}
\caption{Standard parameters for dust grains used in all numerical
calculations. The efficiency factors are from Hanner 1988 for amorphous carbon,
and Ossenkopf, Henning \& Mathis 1992 for (the ``warm'' version of) silicate
grains. The grain size $a$ and sublimation temperature \Tc\ are assumed.  The
lower part lists derived quantities: \QV\ is the efficiency factor for
absorption at visual; \Qast\ is the Planck average at the stellar temperature
of the efficiency coefficient for radiation pressure (equation \ref{q*});
\PsiO\ is defined in equation \ref{Psi0}.}
\smallskip

\end{center}
\end{table}
%%%%%%%%%%%%%%%%%%%%%%%%%%%%%%%%%%%%%%%%%%%%%%%%%%%%%%%%%%%%%%%%%%%%%%%%%%%%%%

\subsection{Dynamics}

We start with a discussion of the three force components controlling the
supersonic phase, identifying in each case the characteristic velocity scale
and the dimensionless profile of its radial variation. These physical processes
have been discussed extensively in the literature, most recently by NE and HTT.
We repeat the essential ingredients to establish the proper formalism and
examine the various assumptions underlying the theory. We also offer a slightly
improved expression for the dust drift velocity, accounting for the subsonic
regime.

\subsubsection{Radiation Pressure}

The radiation pressure force per unit volume is
\eq{
  {\cal F}\sub{rad} = {1\over c}\,\nd\,\upi a^2
            \!\int\!\Q F_\lambda\,d\lambda.
}
Here $Q\sub{pr}$ is the radiation pressure efficiency ($= Q\sub{abs} +
Q\sub{sca}$, assuming isotropic scattering) and $F_\lambda$ is the local
radiative flux, comprised of the attenuated-stellar and diffuse contributions;
note that the diffuse flux vanishes at $r_1$ (IE97). The spectral matching
between $F_\lambda$ and the dust opacity varies in the wind because of the
reddening of the radiation. This variation is conveniently described by the
following radial profile, normalized to unity at $y$ = 1:
\eq{\label{phi}
 \phi(y) = {1\over\Qast}\int \Q{F_\lambda(y) \over F(y)}\, d\lambda.
}
Here $F = \int F_\lambda\,d\lambda$ is the bolometric flux and
\eq{\label{q*}
 \Qast = \int\Q\,
             {F_\lambda(1)\over F(1)}\ d\lambda
       = {\pi\over\sigma T_\ast^4}
         \int\Q B_\lambda(T_\ast)d\lambda,
}
where $B_\lambda$ is the Planck function.  Table 1 lists the values of \Qast\
for our standard grains and $T_\ast$ = 2500 K; the dependence on $T_\ast$ is
insignificant when this quantity is varied within its physical range. Because
of the reddening of the radiation, the spectral matching tends to decrease with
radial distance so that $\phi(y) \le 1$.

The velocity scale associated with the radiation pressure force is defined via
$\vpp = 2\,r_1{\cal F}\sub{rad}(r_1)/\rho(r_1)$. Introduce
\eq{\label{sg}
  \sg = \upi a^2 {\nd\over\nH}\Big|\sub c = \E{-22}\ss\ \hbox{cm}^{2},
}
the cross-section area per gas particle upon dust condensation. Then
\eq{\label{vp}
 \vp  = \left(\Qast\sg L\over2\pi\mp c r_1\right)^{\!\!1/2}
      = 111\ \kms \left(\Qast\ss L_4\over r_{1,14}\right)^{\!\!1/2}\,,
}
where $r_{1,14} = r_1/\E{14}$ cm. Ignoring momentarily the drag and gravity
effects, the radiative force gives the equation of motion
\eq{
 {dv^2\over dy} = {\vpp\over y^2}\,\phi(y).
}
When reddening is neglected too, $\phi$ = 1 and the solution is simply
\eq{\label{GS}
    v^2 = v^2\sub T + \vpp\left(1 - {1\over y}\right).
}
Here we take as the starting point for the outflow velocity the isothermal
sound speed
\eq{\label{vT}
    \vT = \left(k\Tk\over m\sub{H_2}\right)^{1/2}
        = 2.03\,T^{1/2}\sub{k3}\ \kms
}
where \Tk\ = \Tkt\x1000 K is the kinetic temperature at $y$ = 1 (not
necessarily equal to the dust temperature at that point). In this
approximation, first derived in GS, the outflow final velocity is $(v^2\sub T +
\vpp)^{1/2} \simeq \vp$, since \vT\ is usually negligible. Typical values for
the free parameters produce a velocity scale \vp\ considerably higher than
observed outflow velocities, a problem noted by Castor (1981) as a serious
shortcoming of dust-driven wind models. The effects of reddening, drift and
gravity must supplement radiation pressure for a viable explanation.

\subsubsection{Drag}

Collisional coupling accelerates the gas particles and decelerates the dust.
Gilman (1972) has shown that the dust-gas relative velocity reaches steady
state within a distance $\ell \ll r_1$ and that the dust then fully mediates to
the gas the radiation pressure force. In steady-state drift the drag force can
be eliminated and the separate equations of motion for the dust and gas
combined into a single equation. HTT provide a useful discussion of this
approach, which is the one taken here. Still, the dust drift has an important
effect because the radiative acceleration is proportional to $\nd/\nH$, and
separate mass conservation for the dust and the gas implies $\nd/\nH \propto
v/\vd$; in spite of our assumption of prompt dust formation and no further
formation or destruction, the dust abundance varies in the shell because of the
difference between the dust and gas velocities.

In appendix \ref{app:drift} we derive a simple expression for the drift
velocity, including both the subsonic and supersonic regimes. The drift effect
introduces the independent velocity scale
\eq{\label{vm}
 \vm = {\Qast L \over \Mdot c} = 203\ \kms\ {\Qast L_4 \over\Msix}
}
and the steady-state drift velocity is
\eq{
    \vrel = {\vm v\,\phi\over\vT + \sqrt{\vm v\,\phi}}
}
(equation \ref{vdrift}). The significance of subsonic drift is rapidly
diminished with radial distance because $v$ is increasing as the gas
accelerates while \vT\ is decreasing as it cools down. The radial variation of
\vT\ requires the gas temperature profile, a quantity that does not impact any
other aspect of the flow and whose calculation contains large uncertainties.
We avoid these uncertainties and use instead the initial \vT\ throughout the
outflow. This slightly overestimates the overall impact of subsonic drift,
producing a negligible error in an effect that is small to begin with.

The dust velocity is $\vd = v + \vrel$, therefore $\nd/\nH$ varies in
proportion to the dimensionless drift profile
\eq{\label{zeta}
     \zeta(y) = {v\over\vd}
         = {\vT + \sqrt{\vm v\,\phi}\over\vT + \vm\phi + \sqrt{\vm v\,\phi}}\,.
}
Since $\pi a^2\nd(y)/\nH(y) = \sg\zeta(y)$ for $y \ge 1$, the force equation
including the drift effect is
\eq{\label{force a}
  {dv^2\over dy} = {\vpp\over y^2}\,\phi(y)\zeta(y)\,.
}
Because of the drift, at $y = 1$ the radiative force is reduced by a factor
$\zeta(1)$, a significant reduction when $\vm \gg \vT$. The reason is that the
dust particles are produced at the velocity \vT\ with a certain abundance and
the drift immediately dilutes that abundance within a distance $\ell \ll r_1$
so that \nd/\nH\ is diminished already at $y = 1$. In calculations that keep
track separately of the dust and the gas, the two species start with the same
velocity and this dilution is generated automatically (cf NE, HTT). Here we
solve for only one component, the two species start with different velocities
at $y = 1$ and the initial dilution must be inserted explicitly.

It is important to note that $\zeta$ is a monotonically increasing function of
$y$ (see figure \ref{fig:radial} below). Therefore, the dust abundance is the
smallest at $y = 1$ and increases from this minimum during the outflow. At the
wind outer regions, \nd/\nH\ exceeds its initial value by the factor
$\zeta(\infty)/\zeta(1)$.

\subsubsection{Gravity}

The gravitational pull introduces an independent velocity scale, the escape
velocity at $r_1$
\eq{\label{vg}
     \vg  = \left(2GM\over r_1\right)^{\!\!1/2}
          = 15.2\ \kms \left(M_0\over r_{1,14}\right)^{\!\!1/2}\,,
}
which is frequently entered in terms of the dimensionless ratio
\eq{\label{Gamma}
     \Gamma = {\cal F\sub{rad}\over \cal F\sub{grav}}\Bigm|_{r_1}
            = {\vpp\over\vgg} = {\Qast\sg L\over4\pi GM\mp c}
            = 45.8\,\Qast\ss{L_4\over M_0}\,.
}
Gravity does not introduce any radial profile beyond its $y^{-2}$ variation.
Adding the gravitational effects, the equation of motion becomes
\eq{\label{force - full}
 {dv^2\over dy} = {\vpp\over y^2}\left[\phi(y)\zeta(y) - {1\over\Gamma}\right].
}
In the limit of negligible drift and reddening ($\zeta = \phi = 1$), the
gravitational pull reduces the wind terminal velocity from \vp\ to $\vp(1 -
1/\Gamma)^{1/2}$, typically only \about\ 1\% effect.

This is the complete form of the equation of motion, including all the
dynamical processes in the wind. The outflow is fully specified by the four
independent velocity scales \vp, \vm, \vg\ and \vT, which is also the initial
velocity $v(y = 1)$, and the reddening profile $\phi$. This profile varies with
the overall optical depth and is determined from an independent equation, the
equation of radiative transfer.

\subsection{Radiative Transfer}

Because of the spherical symmetry, the radiative transfer equation requires as
input only two additional quantities (IE97). One is the overall optical depth
along a radial ray at one wavelength, which we take as visual
\eq{
    \tV = r_1\sigma\sub V \int\nd dy;
}
the optical depth at every other wavelength is simply $\tV Q_\lambda/\QV$,
where $Q_\lambda$ and \QV\ are the absorption efficiencies at $\lambda$ and
visual, respectively. The other required input is the normalized radial profile
of the dust density distribution
\eq{\label{eta_def}
    \eta(y) = {\nd(y)\over\DS\int_1^{\infty}\!\!\!\!\nd dy}\,.
}
Given these two quantities, the intensity $I_\lambda(y,\beta)$ of radiation
propagating at angle $\beta$ to the radius vector at distance $y$ can be
obtained from
\eq{\label{rad-tran}
 {dI_\lambda(y,\beta) \over d\tau_\lambda(y,\beta)} = S_\lambda - I_\lambda\,,
}
where
\eq{
    \tau_\lambda(y,\beta) = \tV{Q_\lambda\over\QV}
              \int_0^{y\cos\beta}\!\!\!\!\!
              \eta\left(\sqrt{z^2 + y^2\sin^2\beta}\right) dz.
}
In general, $\eta$ and \tV\ must be specified as independent input. Instead,
here they are fully specified by the dynamics problem. With the assumption of
no grain growth or sputtering after its prompt formation, mass conservation for
the dust gives $\nd \propto 1/r^2\vd \propto \zeta/y^2v$ so that
\eq{\label{eta0}
  \eta(y) = {\zeta(y)\over y^2 v(y)}
       \left(\int_1^\infty\!\!\! {\zeta\over y^2 v}\, dy\right)^{\!\!\!-1}.
}
Therefore $\eta$ is not an independent input property, instead it is
determined by the solution of the equation of motion.  And straightforward
manipulations show that
\eq{
    \tV = \QV{\vpp\over2\vm}\int_1^\infty\!\!\!{\zeta\over y^2v}\,dy,
}
so \tV, too, does not require the introduction of any additional independent
input. The radiative transfer problem is fully specified by the parameters
already introduced to formulate the equation of motion; it requires no
additional input.

This completes the formulation of the dusty wind problem. The dynamics and
radiative transfer problems are coupled through the reddening profile $\phi$.
The solution of the equation of motion obtained with a certain such profile
determines a dust distribution $\eta$ and an optical depth \tV. The solution of
the radiative transfer equation obtained with these $\eta$ and \tV\ as input
properties must reproduce the reddening profile $\phi$ that was used in the
equation of motion to derive them.

\subsection{Scaling}

Given grain properties and the spectral profile of the stellar radiation, the
wind problem is fully specified by the four independent velocities \vp, \vm,
\vg\ and \vT, which together form the complete input of the problem. Since an
arbitrary velocity magnitude can always be scaled out, the mathematical model
can be described in a dimensionless form which includes only three parameters.
We choose \vm\ for this purpose because it results in a particularly simple
equation of motion. Introduce
\eq{\label{w-P-theta}
 w = v/\vm, \qquad \theta = \vT/\vm, \qquad P = \vp/\vm\,.
}
The parameter $P$ characterizes the ratio of radiation pressure to drift
effects. It is very large when the drift becomes negligible and goes to zero
when the drift dominates. The equation of motion (\ref{force - full}) becomes
\eq{\label{equation}
 {dw^2\over dy} =
   {P^2\over y^2}\left(\phi\zeta - {1\over\Gamma}\right),
 \quad \hbox{where}\
 \zeta =  {\theta + \sqrt{w\phi}\over\theta + \phi + \sqrt{w\phi}}\,
}
and where $\phi$ is determined from the solution of the radiative transfer
equation (\ref{rad-tran}) in which the dust distribution and optical depth are
\eqarray{\label{eta_tV}
  \eta(y) &=& {\zeta(y)\over y^2 w(y)}
       \left(\int_1^\infty\!\!\! {\zeta\over y^2 w}\, dy\right)^{\!\!\!-1},
       \non
    \tV &=& \case1/2\,\QV P^2\!\!\int_1^\infty\!\!\!{\zeta \over y^2w}\ dy\,.
}
In this new form, the mathematical model is fully specified by the three
dimensionless parameters $P$, $\Gamma$ and $\theta$, where the latter is also
the initial value $\theta = w(y = 1)$. The velocity must rise at the origin,
and since $\phi(1) = 1$ the parameters are subject to the constraint
\eq{\label{liftoff}
    \Gamma\zeta(1) > 1, \qquad \hbox{i.e.}\quad
    (\Gamma - 1)(\theta + \sqrt{\theta}) > 1.
}
This liftoff condition ensures that radiation pressure, with proper accounting
for the dust drift, can overcome the initial gravitational pull. It can be
viewed as either a lower limit on $\Gamma$ for a given $\theta$ or, given
$\Gamma$, a lower limit on $\theta$.

The dusty wind problem has been transformed into a set of two general
mathematical equations --- the equation of motion (\ref{equation}) and the
radiative transfer equation (\ref{rad-tran}) with the input properties from
equation (\ref{eta_tV}). Given the spectral shapes of the stellar radiation
and the dust absorption coefficient, these equations are fully prescribed by
$P$, $\Gamma$ and $\theta$. Both the radiative transfer and the dynamics
equations are solved without any reference to \vm\ or any other velocity
scale. The velocity scale is an extraneous parameter, entirely arbitrary as
far as the solution is concerned. {\em The wind radiative emission and the
shape of its velocity profile are both independent of the actual magnitude of
the velocity.} In the following discussion we show that the shape of $w(y)$
turns out to be a nearly universal function, independent of input parameters
within their relevant range, and that the final velocity $\wf = w(y
\to\infty)$ is for all practical purposes only a function of $P$.

\section{SOLUTIONS}

The complete solution involves two elements --- dynamics and radiative
transfer. The impact of the radiative transfer on the dynamics can be
conveniently expressed in terms of the quantity $\Phi$, defined via
\eq{\label{Phi}
    {\wf\over\Phi} = \int_{\theta}^{\wf}{dw\over\phi}\,.
}
That is, $\Phi$ is the velocity-weighted harmonic average of the reddening
profile $\phi$. Combining equations \ref{equation} and \ref{eta_tV}, it is easy
to show that
\eq{\label{tV_work}
    {\tV\over\QV} = {\wf\over\Phi}
    + {P^2\over2\Gamma}\int_1^\infty{dy\over y^2w\phi}\,.
}
Therefore, when gravity is negligible ($\Gamma \gg 1$) the optical depth and
the terminal velocity obey the simple relation
\eq{\label{wf-tauV}
    \wf = \Phi{\tV\over\QV}\,.
}
In optically thin winds $\Phi = 1$ and \wf\ = \tV/\QV. As reddening increases
$\Phi$ decreases, the final velocity is affected by the radiative transfer and
numerical computations are necessary.

We obtain numerical solutions for arbitrary optical depths with the code DUSTY
(Ivezi\'c, Nenkova \& Elitzur, 1999).  DUSTY solves the dusty wind problem
through a full treatment of the radiation field, including scattering,
absorption and emission by the dust, coupled to the hydrodynamics problem
formulated above. Appendix \ref{numerics} provides a description of the
numerical procedure.  We present now the solutions, starting with the optically
thin regime where we found the exact analytic solution.

\subsection{Negligible Reddening}
\label{sec:thin}

When reddening is negligible, the flux spectral shape does not vary and $\phi =
1$ (see eq. \ref{phi}). This is the situation in optically thin winds, and the
detailed results presented in the next section show this to be the case when
$\tV \la 1$.\footnote{Another limit that in principle could give $\phi = 1$
involves grains so large that $Q_\lambda$ is constant for all relevant
wavelengths. This requires grain sizes in excess of \about\ 10 \mic, and seems
of little relevance.} Even though the optical depth at wavelengths shorter
than visual already exceeds unity when \tV\ = 1 and the emerging spectrum is
affected by radiative transfer, the impact on the dynamics is minimal. The
reason is that $\phi$ involves a spectral average, and the stellar radiation
peaks at longer wavelengths.

The wind problem with $\phi = 1$ was called the ``first simplified model" by
HTT, who solved it numerically. Here we present the complete analytic solution
for this case. Since $\phi$ is known, the equation of motion decouples from the
radiative transfer problem and can be considered independently. Introduce
\eq{
    \delta = {1\over\Gamma - 1},
}
then equation \ref{equation} can be cast in the form
\eq{
    \left(1 + {1 + \delta\over\sqrt{w} + \theta - \delta}\right){dw^2\over dy}
        = {P^2\over 1 + \delta}\,{1\over y^2}
}
so that
\eq{\label{solution}
  w^2 - \theta^2 + 4(1 + \delta)\!\!\!\int\limits_{\sqrt{\theta}}^{\sqrt{w}}
 \!\!{x^3dx\over x + \theta - \delta} =
 {P^2\over 1 + \delta}\left(1 - {1\over y}\right).
}
The liftoff constraint (\ref{liftoff}) ensures that the denominator of the
integrand never vanishes in physical solutions; it starts positive at the
lower boundary and increases as $x$ increases. The integration is standard, the
result is
\eqarray{\label{CompleteSolution}
    {P^2\over1 + \delta}\bigg(1 - {1\over y}\bigg) &= &w^2 - \theta^2
    + 4(1 + \delta)\bigg[\case1/3(w^{3/2} - \theta^{3/2})       \non
     &&+\ \case1/2(\delta - \theta)(w - \theta)                 \non
     &&+\ (\delta - \theta)^2(w^{1/2} - \theta^{1/2})           \non
     &&+\ (\delta - \theta)^3
   \ln{\sqrt{w} + \theta - \delta\over\sqrt{\theta} + \theta - \delta} \bigg].
}
This is the complete solution for all dusty winds with $\tV < 1$, fully
incorporating the effects of gravity and dust drift. Radiation reddening, which
takes effect when $\tV > 1$, is the only ingredient missing from this analytic
solution and preventing it from applicability for all winds. The GS solution
(equation \ref{GS}), the only previous analytic result, which neglected the
effects of gravity and drift in addition to reddening, can be recovered
inserting $\delta$ = 0 and $P \gg 1$.
%%%%%%%%%%%%%%%%%%% Thin Winds Solution %%%%%%%%%%%%%%%%%%%%%%%%%%%%%
\begin{figure*}
\begin{minipage}{\textwidth}
\centering \leavevmode
 \psfig{file=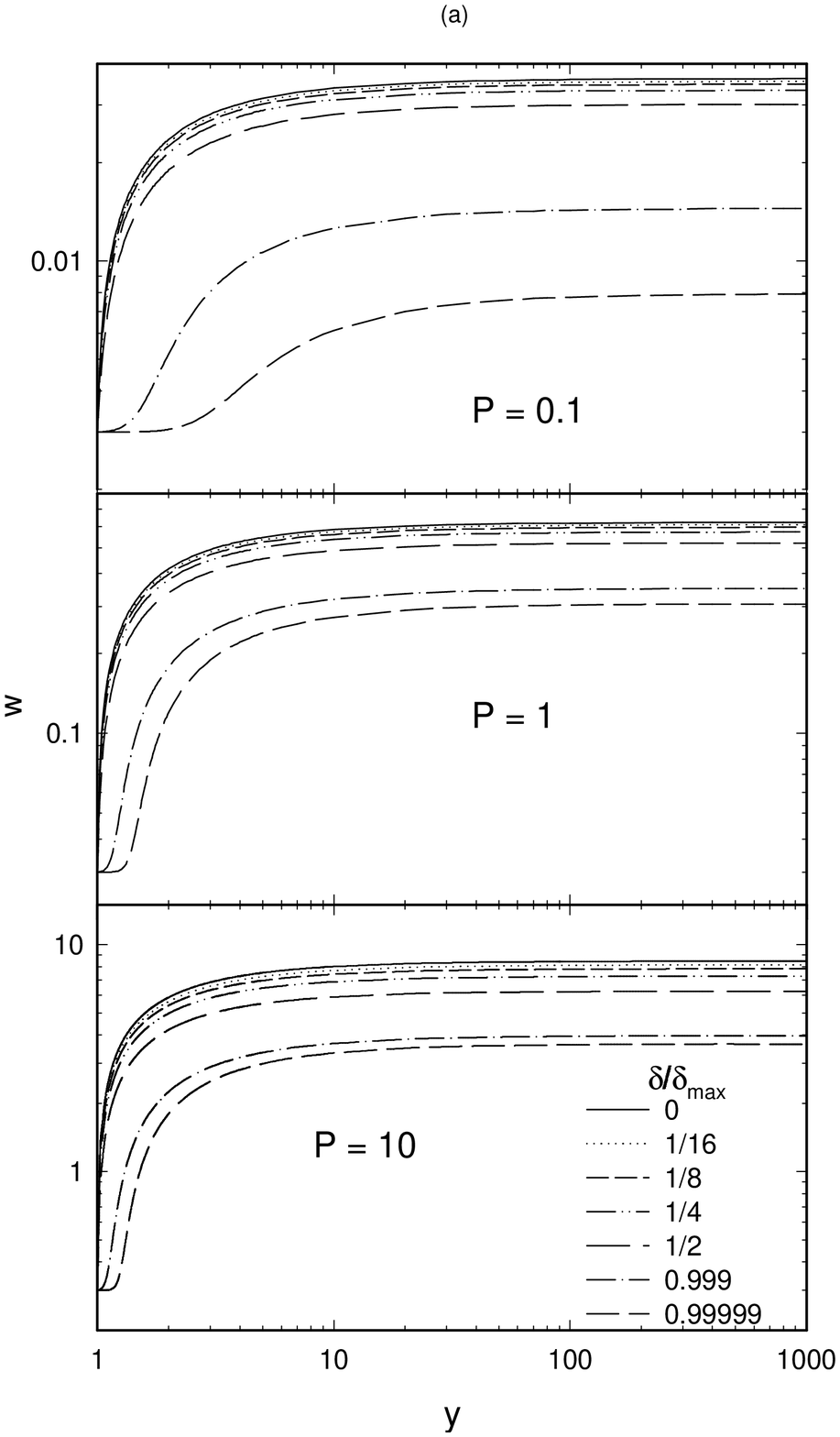,height=0.8\hsize,clip=} \hfil
 \psfig{file=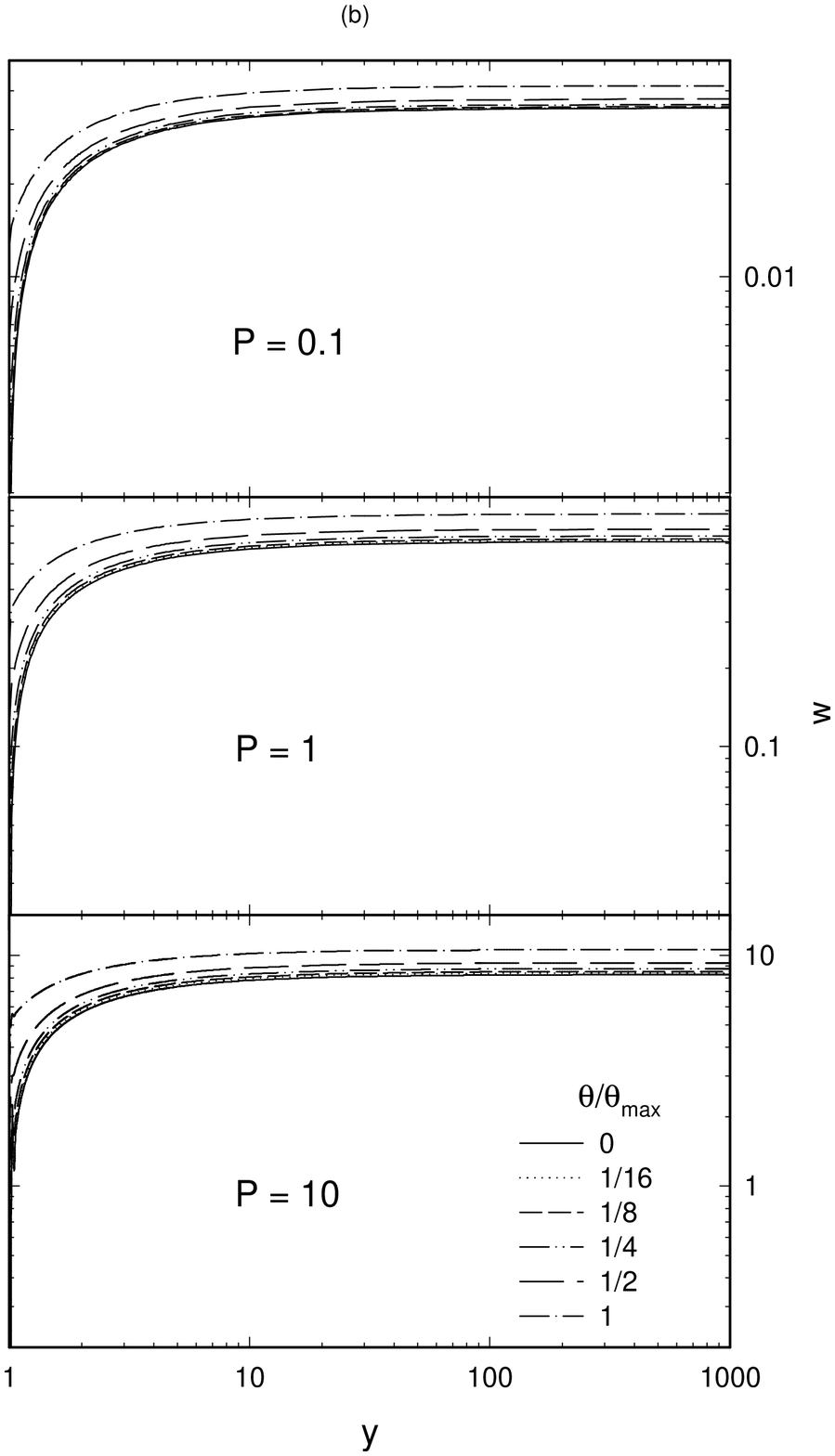,height=0.8\hsize,clip=}
\caption{The complete solution (equation \ref{CompleteSolution}) for the
velocity profiles of optically thin winds. (a) In each panel $P$ has the
indicated value and the initial velocity is $\theta = 0.03P$. The solutions are
plotted for various values of $\delta$, marked as fractions of the
gravitational-quenching value \dmax. Note the virtual $\delta$-independence,
except for the solutions asymptotically close to \dmax.  (b) Same as (a), only
now $\theta$ varies and $\delta$ is fixed at 0. The values of $\theta$ are
listed as fractions of $\theta\sub{max}$, the initial velocity that yields the
smallest meaningful velocity increase $\wf/\theta = \vf/\vT = 3$. Note the
complete $\theta$-independence, except for the solution with $\theta =
\theta\sub{max}$.} \label{fig:thin}
\end{minipage}
\end{figure*}
%%%%%%%%%%%%%%%%%%%%%%%%%%%%%%%%%%%%%%%%%%%%%%%%%%%%%%%%%%%%%%%%%%%%%%%%%%%%

The solutions are meaningful only when describing winds in which $\vf/\vT$ is
at least \about\ 3; otherwise, the effects of gas pressure, which were
neglected here, become significant and equation \ref{equation} loses its
physical relevance. Therefore, we require $\vf/\vT = \wf/\theta > 3$. In
addition, the liftoff condition implies that $\delta < \dmax$ where
\eq{\label{dmax}
    \dmax = \sqrt{\theta} + \theta .
}
Figure \ref{fig:thin} displays the dimensionless velocity profile as a function
of distance $y$ for a range of representative values of the three free
parameters and demonstrates a remarkable property: Except for their role in
determining the boundary of allowed phase space and controlling the wind
properties near that boundary, $\delta$ and $\theta$ hardly matter; away from
the boundary, the solution is controlled almost exclusively by the single
parameter $P$. Part (a) of the figure shows the effect of varying $\delta$ when
$P$ and $\theta$ are held fixed. Each panel presents a representative value of
$P$ with a wind initial velocity $\theta = 0.03P$; this choice of $\theta$
ensures that the wind velocity increase by a factor of \about\ 10--20 in each
cases. The displayed values of $\delta$ cover the entire physical range for
this parameter, from 0 to just below the singularity at \dmax. In each panel,
the solutions for $\delta/\dmax \le \case1/2$ are rather similar to each other,
those with $\delta/\dmax \le \case1/4$ hardly distinguishable. All display a
similar rapid rise within $y \la 10$ toward a final, nearly the same velocity.
The plots for $\delta/\dmax \ge 0.999$ stand out and show the quenching effect
of gravity, which is discussed below (sec. \ref{Quenching}).

The effect of $\theta$ on the solution when $\delta$ and $P$ are held fixed are
similarly shown in part (b) of the figure. In these panels $\delta$ = 0. This
value was chosen because it represents faithfully most non-quenched solutions
while allowing the wind to start with arbitrarily small initial velocity, even
zero. The values of $\theta$ for the displayed solutions are listed as
fractions of $\theta\sub{max}$, the initial velocity that leads to $\wf/\theta$
= 3 for the listed $P$. As is evident from the plots, the initial velocity is
largely irrelevant. Starting from $\theta = 0$ or as much as
$\case1/2\theta\sub{max}$ yields practically the same results. The only
discernible difference occurs at $\theta\sub{max}$, where the profile deviates
from the common shape by no more than \about\ 10\%.

The analytic solution defines $w$ as an implicit function of $y$.  In appendix
\ref{analytic} we derive explicit analytic expressions for $w$ that provide
adequate approximations in all regions of interest. An inspection of the
solution shows that $P$ = 16/9 is the transition between the drift-dominated
regime at small $P$ and negligible drift at large $P$. The velocity profiles in
the two are
\eq{\label{profile1}
    w = \wf\left(1 - {1\over y}\right)^k,     \qquad
    k = \cases{2/3 & for $P < 16/9$ \cr
                   &            \cr
               1/2 & for $P > 16/9$}
}
The expressions for \wf\ in the two regimes can be combined into the single
form
\eq{\label{wf0}
 \wf = P\left(P\over P + \case{16}/9\right)^{\!\!1/3}\,.
}
The optical depth, needed for the radiative transfer problem which is solved
independently in this case, is simply \tV\ = \QV\wf\ (see eq.\ \ref{wf-tauV}).
Appendix \ref{analytic} provides the $\delta$- and $\theta$-corrections to
these results, which show that the dependence on $\theta$ is confined to the
very origin of the wind, $y - 1 \la \theta/\wf$.

The last two expressions reproduce to within 30\% the plots for all the
non-quenched winds displayed in figure \ref{fig:thin}. They amplify our
conclusion about the negligible role of both $\delta$ and $\theta$ away from
the physical boundary and point to another remarkable property of the solution.
Although the final velocity strongly depends on $P$, the shape of the velocity
profile $w/\wf$ does not. The only reference to $P$ in this profile is in
determining a transition to a slightly steeper profile shape at large $P$.  The
profile itself is independent of $P$ in either regime and furthermore, the
difference between the two shapes is not that large. These properties are
evident from figure \ref{fig:thin}. Other than the numerical values on the
vertical axes, it is hard to tell apart the plots in the different panels.

These results show that among the three independent input parameters, $P$ is
the only one to have a significant effect on any property of interest.
Furthermore, even $P$ affects mostly just the final velocity, it does not have a
discernible effect on the shape of the velocity profile; {\em all optically
thin winds share universal velocity and dust density profiles}.

\subsection{Reddening Effects}

The results for optically thin winds carry over to arbitrary optical depths,
with profound implications for all winds. Reddening effects are controlled by
the wind optical depth and density profile. For a given pair of $\delta$ and
$\theta$, consider a value of $P$ sufficiently small that $\tV \ll 1$ so that
reddening can be neglected; such a choice of $P$ is virtually always possible.
The velocity profile and \tV\ are then uniquely determined by $P$ (equations
\ref{profile1} and \ref{wf0}) and so is the radiative transfer problem. When
$P$ increases, \tV\ increases too. The results of Appendix \ref{analytic} show
that
\eq{\label{P1}
    P = {1\over\QV}\left(1 + \case4/3Q^{1/2}\sub{V}\right)^{1/2}
}
yields \tV\ = 1; with our standard grains, the corresponding values are $P =
0.73$ for carbon and $P = 1.35$ for silicates.  As $P$ increases further the
optical depth increases too and with it the impact of reddening, and that
impact too is controlled exclusively by $P$. Therefore, the parameter $P$
controls all aspects of the problem. As concluded in IE95, away from its
boundary and for most of phase space {\em the dusty wind problem is controlled
by a single free parameter}. Since $P$, \tV\ and \wf\ are uniquely related to
each other, either one can serve as that free parameter.

%%%%%%%%%%%%%%%%%%% Radial Profiles %%%%%%%%%%%%%%%%%%%%%%%%%%%%%
\begin{figure}
 \centering \leavevmode \psfig{file=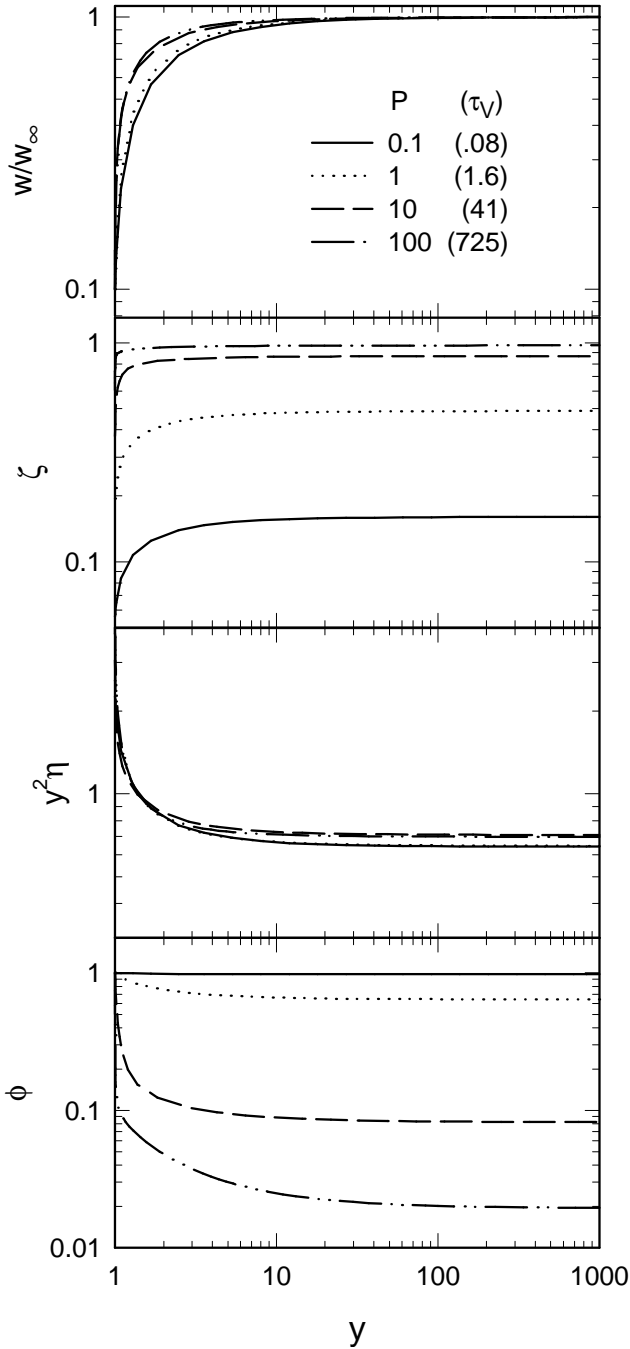,width=\figsize,clip=}
\caption{Radial profiles for quantities of interest at various values of $P$,
as marked. The corresponding optical depths are shown in parentheses. All
models have amorphous carbon grains, $\delta$ = 0 and $\wf/\theta$ = 10.}
\label{fig:radial}
\end{figure}
%%%%%%%%%%%%%%%%%%%%%%%%%%%%%%%%%%%%%%%%%%%%%%%%%%%%%%%%%%%%%%%%%%%%%%%%%%%%

Figure \ref{fig:radial} shows the radial variation of profiles of interest for
various values of $P$. The models for $P$ = 0.1, 1 and 10 repeat those
presented in figure \ref{fig:thin} but fully incorporate radiative transfer.
Comparison of the corresponding plots in the two figures illustrates the impact
of reddening. The bottom panel of figure \ref{fig:radial} shows the reddening
profile $\phi$. Most of the reddening occurs close to wind origin. Reddening
becomes significant at $P = 1$, and substantial at $P$ = 10 and 100. Still it
has only a minimal impact on the velocity profile; as is evident from the top
panel, the dependence of the profile shape on $P$ remains weak. We find that
equation \ref{profile1} remains an excellent approximation under all
circumstances, the effect of reddening merely decreases the exponent $k$ from
0.5 to 0.4 at large $P$.  The plots of $\zeta$ show that most of the drift
variation, too, occurs close to the origin. At $y \ga 2$ the dust and gas
velocities maintain a constant ratio, which can be substantial when $P$ is
small; at $P = 0.1$ the final velocity of the dust grains exceeds that of the
gas particles by more than factor 6. The velocity and drift profiles combine to
determine the dust density profile $\eta$ (see eq. \ref{eta_tV}). The figure
presents the function $y^2\eta$, removing the common radial divergence factor
$1/y^2$.

%%%%%%%%%%%%%%%%%%% P-Profiles %%%%%%%%%%%%%%%%%%%%%%%%%%%%%
\begin{figure}
 \centering \leavevmode \psfig{file=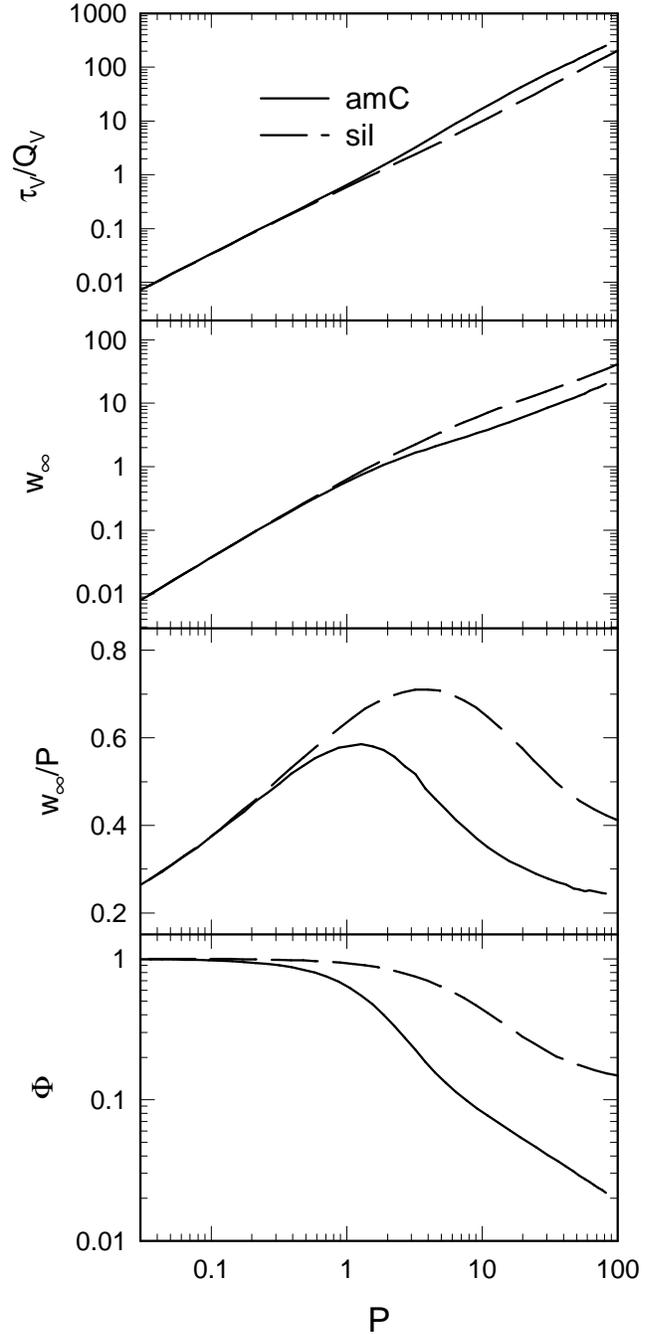,width=\figsize,clip=}
\caption{Variation with P of various quantities of interest.}
\label{fig:P-profiles}
\end{figure}
%%%%%%%%%%%%%%%%%%%%%%%%%%%%%%%%%%%%%%%%%%%%%%%%%%%%%%%%%%%%%%%%%%%%%%%%%%%%

The plots present the full solutions of the dynamics problem coupled with
radiative transfer for carbon grains. The results for silicate grains are
essentially the same, extending the conclusion reached in the optically thin
limit to all cases: {\em all radiatively driven dusty winds share nearly
identical velocity and dust density profiles.}

Figure \ref{fig:P-profiles} presents the dependence of \tV/\QV\ (= \wf/$\Phi$)
and final velocity \wf\ on $P$.  The bottom panel shows the reddening indicator
$\Phi$. As long as $\Phi = 1$, reddening is negligible, the analytic solution
(\ref{CompleteSolution}) holds and the grain properties are irrelevant. Indeed,
all quantities are the same for carbon and silicate dust in that regime. The
grain composition matters only when reddening affects the dynamics. The figure
shows that $\Phi$ begins to decrease, indicating that reddening becomes
significant, at a value of $P$ corresponding to $\tV \simeq 1$; the actual
value is controlled by \QV\ (equation \ref{P1}), it is larger for silicate
because of its smaller \QV. As noted before, the optical depth at wavelengths
shorter than visual already exceeds unity when \tV\ = 1, but that spectral
region has little impact on the dynamics.

Appendix B shows that the analytic approximation in equation \ref{wf0} can be
extended to include reddening effects,
\eq{\label{wf3}
 \wf = P\sqrt{\Phi}\,\left(P\over P + \case{16}/9\sqrt{\Phi}\right)^{\!\!1/3}.
}
This expression provides an excellent approximation for the numerical results
presented in figure \ref{fig:P-profiles}; the deviations are less than 10\%
for silicate grains, 30\% for carbon. In addition to \wf, the figure displays
also $\wf/P = \vf/\vp$.  As discussed in \S\ref{Theory}, the simplest
application of radiation pressure gives the GS result \vf\ = \vp, i.e. $\wf/P
=1$. The figure shows that the actual solution has a substantially different
behavior --- $\wf/P$ is neither constant, nor does it reach unity. Instead, it
increases in proportion to $P^{1/3}$ at $P \la 1$, because of the drift, and
decreases in proportion to $\Phi^{1/2}$ at $P \ga 1$, because of the reddening.
Its maximum, reached around $P$ \about\ 1, is larger for silicate because it
requires larger $P$ for reddening to become significant. Drift and reddening
play a crucial role in the behavior of $\wf/P$. The maximum reached by this
function translates to a maximum velocity for dusty winds, discussed in the
next section.

\section{OBSERVABLE CORRELATIONS}
\label{sec:physical}

Contact with observations is made by transforming the mathematical problem back
into physical quantities. Except for regions close to the boundary of phase
space (see \S\ref{sec:boundaries}), the wind problem is fully controlled by the
parameter $P$ and its solution determines the corresponding \wf. Both
quantities will now be expressed in terms of physical parameters.

The main scaling variable is
\eq{\label{P}
    P = 0.546\, \Msix\left(\ss\over\Qast\, L_4\, r_{1,14}\right)^{1/2}
}
(see equation \ref{w-P-theta}). This expression does not determine $P$
explicitly because $P$ enters indirectly also on its right hand side.  The dust
condensation condition $T(r_1) = \Tc$ determines $r_1$ as (IE97)
\eq{\label{r1}
 r_1 = \left({L\Psi\over16\upi\sigma T\sub{c}^4}\right)^{1/2}
     = 1.16\x\E{14}{L_4^{1/2}\Psi^{1/2} \over T^2\sub{c3}} \ \hbox{cm.}
}
Here \Tct\ = \Tc/(1000 K) and $\Psi$ is a dimensionless function determined by
the radiative transfer, similar to the reddening profile, and thus dependent on
$P$. Optically thin winds have $\Psi = \PsiO$ where
\eq{\label{Psi0}
    \PsiO = {\QP(T_\ast)\over\QP(\Tc)}\,.
}
Here $\QP(T)$ is the Planck average of the absorption efficiency, similar to
the average of the radiation pressure efficiency that defines \Qast\ (equation
\ref{q*}); Table 1 lists \PsiO\ for our standard grains. As $P$ increases, the
wind becomes optically thick and $\Psi$ increases too. In IE97 we present the
variation of $\Psi/\PsiO$ with optical depth and show that it is well
approximated by the analytic expression
\eq{\label{Psi}
    {\Psi\over\PsiO} = 1 + 3{\tV\over\QV}\,\QP(\Tc)\fbar
}
where
\[
    \fbar = \int\phi(y)\eta(y){dy\over y^2}
\]
(see figure 1 and equation B7 in IE97)\footnote{Since scattering is neglected
in the IE97 analytic solution, the difference between \Qast\ and $\QP(T_\ast)$
is ignored in the approximations here (but not in the numerical
calculations).}. Just as $\Phi$ is the velocity-weighted harmonic average of
the reddening profile, \fbar\ is its standard average, weighted by $\eta/y^2$.

For most of the relevant region of phase space, $P$ fully controls the solution
of the dusty wind problem. Since $\Psi$ is part of that solution, it too
depends only on $P$. Therefore, the combination of equations \ref{P} and
\ref{r1} results in
\eq{\label{R1}
    {\Msix\over L_4^{3/4}} = 1.98\,{Q_{\ast}^{1/2}\over\sigma_{22}^{1/2}\Tct}
                        \, P\Psi^{1/4},
}
a one-to-one correspondence between $\Mdot/L^{3/4}$ and $P$.  This
correspondence involves the function $P\Psi^{1/4}$, which is fully determined
from the solution of the mathematical wind problem. The proportionality
constant involves the individual grain properties \Qast\ and \Tc\ which are
part of the problem specification, and \ss\ which is not. The dust abundance,
necessary for specifying \ss, has no effect on the wind solution; only the
proportionality constant is modified when this abundance is varied, the
function $P\Psi^{1/4}$ remains the same.

This result shows that for any dusty wind, the combination $\Mdot/L^{3/4}$ can
be determined directly from the shape of the spectral energy distribution: the
wind IR signature is fully controlled by the parameter $P$, therefore comparing
the observations with a bank of solutions determines the value of $P$ and
equation \ref{R1} fixes $\Mdot/L^{3/4}$. We may expect all C-rich stars to have
dust with roughly similar abundance and individual grain properties, so that
they have the same proportionality constant and functional dependence
$\Psi(P)$; likewise for O-rich stars. In that case, each family defines a
unique correspondence between $\Mdot/L^{3/4}$ and $P$. And since the parameter
$P$ fully controls the IR signature, {\em stars that have different \Mdot\ and
$L$ but the same $\Mdot/L^{3/4}$ are indistinguishable by their IR emission}
because they also have the same $P$.

The solution of the wind problem also determines $\wf$, another unique function
of $P$. Therefore, the relation $\vf/\vm = \wf$ gives another one-to-one
correspondence between $P$ and an independent combination of physical
quantities
\eq{\label{R2}
    {\Mdot\,\vf\over L} = {1\over c}\,\Qast\,\wf.
}
Since this correspondence does not involve \ss, it is independent of the dust
abundance and depends only on individual grain properties. {\em Systems with
different \Mdot, $L$ and \vf\ but the same $\Mdot/L^{3/4}$ necessarily have
also the same $\Mdot\vf/L$} because each combination uniquely determines the
parameter $P$ of the system.

The complete description of the wind problem involves four velocity scales, two
of which (\vT\ and \vg) do not count whenever $\Gamma$ and $\theta$ can be
ignored. For most of phase space the problem is fully specified by the
velocities \vp\ and \vm, and its solution determines \vf. The two ratios formed
out of these three velocities are all the independent dimensionless
combinations of physical parameters to determine the mathematical wind model
and its solution. Since this solution fully specifies the wind IR signature,
{\em $\Mdot/L^{3/4}$ and $\Mdot\vf/L$ are the only combinations of global
parameters that can be determined from IR observations}, even the most detailed
ones. When the velocity is additionally measured in molecular line observations
so that both $P$ and \vf\ are known, the two combinations in equations \ref{R1}
and \ref{R2} can be used to determine also $L$ and \Mdot\ individually. The
relevant correlations are
\eq{\label{R3}
    {\vf\over L^{1/4}} \propto {\wf\over P\Psi^{1/4}}\,, \hspace{0.5in}
    {\vfff\over\Mdot}  \propto {\wfff\over P^4\Psi}\,,
}
whose constants of proportionality can be trivially derived. Only two of the
last four combinations that relate \Mdot, $L$ and \vf\ to the solution are
independent; any two of them can be derived from the other two.

\subsection{Similarity Relations}
\label{scaling}

While $P$ is the natural independent variable of the mathematical problem, its
physical interpretation is only indirect. Modeling of IR observations directly
determines \tV, not $P$. Therefore, we now express all results in terms of \tV\
by performing a straightforward change of variables. The relation between \tV\
and $P$ is non-linear, starting from $P \propto \tau^{3/4}\sub V$ in the
optically-thin regime and switching to a more complex dependence when reddening
becomes significant. We introduce the \tV--$P$ transformation function $\Theta$
through
\eq{\label{Theta0}
    P = {2\over\sqrt{3}}\left(\tV\over\QV\right)^{3/4}\Theta
}
so that $\Theta$ = 1 when $\tV < 1$. At larger optical depths, $\Theta$ can
only be determined from the numerical solution. The analytic expression
\eq{\label{Theta}
    \Theta^2 \simeq \left[1 + \case3/4(\tV/\QV)^{1/2}\right]\Phi\,,
}
obtained from equation \ref{P-tauV}, provides a useful approximation for the
numerical results at all optical depths. With the \tV--$P$ transformation,
equation \ref{R1} becomes
\eq{\label{K1}
  {\Msix\over L_4^{3/4}} = c_1\,\tau^{3/4}\sub VK_1(\tV),
}
where
\[  K_1 = \left(\Psi\over\PsiO\right)^{\!\!1/4}\Theta, \hspace{0.3in}%\qquad
    c_1 = 2.28{Q_{\ast}^{1/2}\Psi_0^{1/4}\over
                 Q^{3/4}\sub V\sigma_{22}^{1/2}\Tct}\,.
\]
From its definition $K_1 = 1$ when $\tV < 1$, therefore $\Msix =
c_1(L_4\tV)^{3/4}$ in all optically thin winds. As the optical depth increases,
$K_1$ introduces the reddening correction to this relation. This correction,
purely a function of \tV, is shown in the top panel of figure
\ref{fig:K-Functions}.

The transformation of the other independent correlation from $P$ to \tV\ is
considerably simpler. Thanks to equation {\ref{wf-tauV}, which holds for all
winds when gravity is negligible, equation \ref{R2} can be cast similarly in
the form
\eq{\label{K2}
  {\Mdot\vf\over L} = {1\over c}\,(\Qast/\QV)\,\tV K_2(\tV),
}
where $K_2 = \Phi$. The function $K_2$, shown in the second panel of figure
\ref{fig:K-Functions}, contains the reddening corrections to the optically thin
correlation, similar to the previous result. Since $K_2 = \Phi$, it is the same
function as shown in figure \ref{fig:P-profiles}, only plotted against the
independent variable \tV\ instead of $P$.

This completes the transformation from $P$ to \tV\ of the two independent
correlations. In both, the luminosity does not enter independently, only
through the product $L\tV$. In the case of equation \ref{K2} this is a direct
consequence of the structure of the equation of motion, which gives $\vf =
\vm\wf = \vm\tV(\Phi/\QV)$. The combination $L\tV$ emerges here as a
``kinematic" result irrespective of the explicit expression for the drift
function. In the case of equation \ref{K1}, on the other hand, this combination
is a direct consequence of the specific functional form of $\zeta$. With these
two results, the correlations in eq.\ \ref{R3} become similarly
\eq{\label{K3-4}
  {v_1\over L_4^{1/4}} = c_3\,\tau^{1/4}\sub VK_3(\tV), \qquad
  {v^3_1\over \Msix}   = c_4\,K_4(\tV)
}
where $K_3 = K_2/K_1$ and $K_4 = K_2^3/K_1^4$. The associated proportionality
constants are
\[  c_3 = 88.9\, \Tct(\Qast\ss)^{1/2}(\QV\PsiO)^{-1/4} \]
\eq{\label{c3-4}
    c_4 = 3.08\x\E5\, T^4\sub{c3}\Qast\ssq\Psi_0^{-1}\,.
}
Since $L$ enters only in the form $L\tV$, the combination that eliminates $L$
eliminates also \tV\ from its optically thin regime. The reddening correction
functions $K_3$ and $K_4$ are shown in the two bottom panels of figure
\ref{fig:K-Functions}.

%%%%%%%%%%%%%%%%%%% K functions %%%%%%%%%%%%%%%%%%%%%%%%%%%%%
\begin{figure}
 \centering \leavevmode \psfig{file=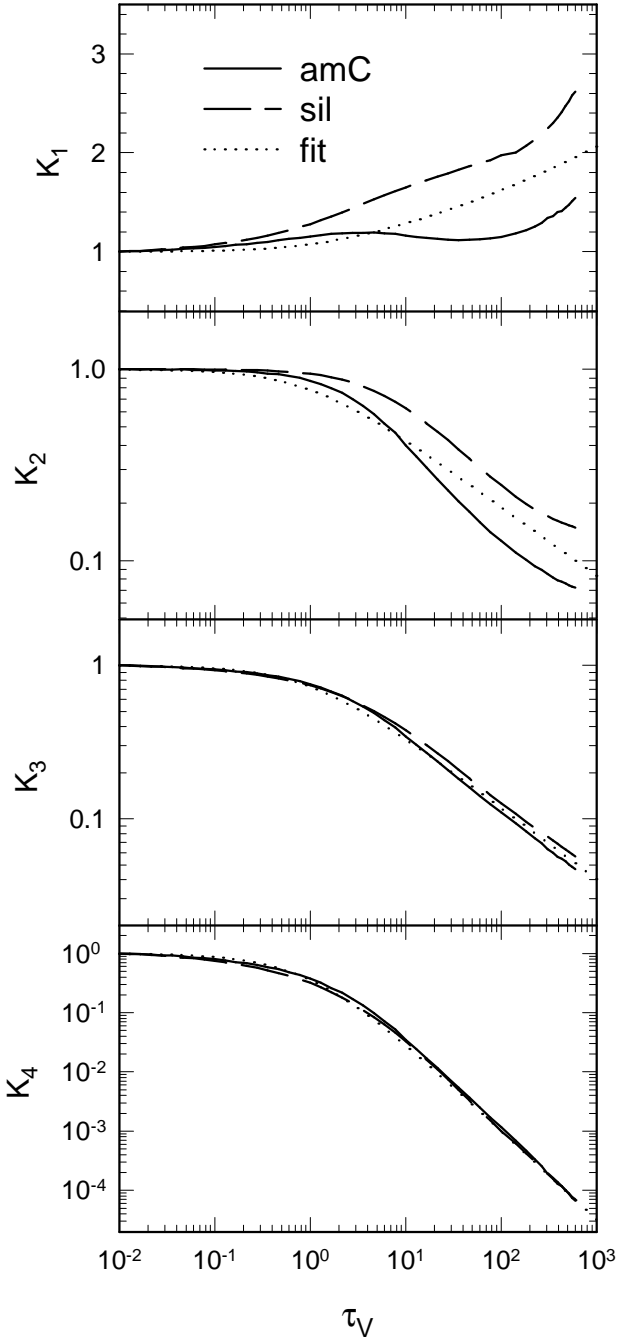,width=\figsize,clip=}
\caption{The reddening corrections in the scaling relations that summarize the
physical contents of the solution (equations \ref{K1}--\ref{K3-4}). Note the
linear scale in the plot of $K_1$. The full and dashed lines show the results
of detailed numerical calculations for carbon and silicate grains,
respectively. The dotted-line in the panel for $K_i$ ($i = 1\dots4$) plots the
function $1/(1 + \tV)^{\alpha_i}$ where $\alpha_1 = -0.105$, $\alpha_2 = 0.36$,
$\alpha_3 = 0.465$ and $\alpha_4 = 1.5$.
} \label{fig:K-Functions}
\end{figure}
%%%%%%%%%%%%%%%%%%%%%%%%%%%%%%%%%%%%%%%%%%%%%%%%%%%%%%%%%%%%%%%%%%%%%%%%%%%%

At small \tV\ the grain properties are irrelevant and the similarity
$K$-functions are unity for both silicate and carbon dust. As \tV\ increases
the curves for the two species diverge, reflecting their different optical
properties. However, figure \ref{fig:K-Functions} shows that the differences
are quite moderate in all four cases. They are most noticeable in $K_1$ and
$K_2$, where the relative differences from the mean never exceed \about\ 25\%.
It is also important to note that the displayed range of \tV\ greatly exceeds
the observed range since cases of $\tV \ga 10$ are rare. The differences are
greatly reduced in the ratio function $K_3$ (= $K_2/K_1$) and practically
disappear in $K_4$. This function is essentially the same (within a few
percent) for both silicate and carbon grains at all optical depths, an
agreement maintained over many orders of magnitude. Evidently, $K_4$ is
independent of chemical composition.

The universality of $K_4$ is no accident. From the analytic approximations for
$\Psi$ (eq.\ \ref{Psi}) and $\Theta$ (eq.\ \ref{Theta}), at large optical
depths $\tau^2\sub V K_4 \propto \Phi/\fbar$; that is, apart from the explicit
dependence on \tV, the variation of $K_4$ comes from the ratio of two averages
of the reddening profile, one ($\Phi$) strongly weighted toward the outer part
of the shell the other (\fbar) toward the inner part. Although $\Phi$ and
\fbar\ are different for grains with different optical properties, their ratio
is controlled by the shapes of the density and velocity profiles --- which are
universal, as shown above. As a result, the functional dependence of $K_4$ on
\tV\ is the same for all grains when $\tV \gg 1$ in addition to $\tV < 1$. This
does not yet guaranty a universal profile because the transition between the
two regimes could depend on the grain parameters. Indeed, the plot of $\Phi$ (=
$K_2$) shows that the large-\tV\ decline of this function is delayed for
silicate in comparison with carbon dust. As explained in the discussion of
figure \ref{fig:P-profiles}, the onset of reddening requires a larger \tV\ for
silicate because of its smaller \QV. But the behavior of the factor $1 +
\case3/4(\tV/\QV)^{1/2}$ in $\Theta$ (eq.\ \ref{Theta}) is precisely the
opposite since its large-\tV\ behavior starts earlier when \QV\ is smaller. The
two effects offset each other, producing a universal shape.

The simple analytic expression $(1 + \tV)^{-1.5}$ fits the function $K_4$ to
within 20\% over a variation range covering more than four orders of magnitude,
providing the nearly perfect fit evident in figure \ref{fig:K-Functions}.
Therefore, dusty winds obey the condition
\eq{\label{C1}
   {v^3_1\over \Msix}\left(1 + \tV\right)^{1.5} = c_4\,.
}
This result provides a complete separation of the system global parameters from
the grain parameters. The combination of \vf, \Mdot\ and \tV\ on the left hand
side is always constant, its magnitude is determined purely by the dust
properties. This expression makes it evident again that the optically thin
regime corresponds to $\tV < 1$.

As is evident from figure \ref{fig:K-Functions}, $K_3$ too is nearly the same
for carbon and silicate grains. The reason is that $K_3 = (K_2K_4)^{1/4}$ and
the only difference between the two species comes from their $K_2$ profiles,
which enter only in the fourth root. The single analytic expression $(1 +
\tV)^{-0.465}$ provides an excellent fit for both silicate and carbon grains,
leading to the independent correlation
\eq{\label{C2}
    v_1 = c_3\,\left(L_4\tV\right)^{1/4}\left(1 + \tV\right)^{-0.465}.
}
This result can explain the narrow range of velocities observed in dusty
winds. The dependence of velocity on \tV\ at fixed luminosity and grain
properties is shown in figure \ref{fig:aux-Functions}. This function reaches a
maximum of 0.73 at \tV\ = 1.3, therefore the largest velocity a dusty wind can
have is
\eq{\label{vmax}
    v\sub{max} = 0.73\,c_3\,L_4^{1/4}\quad \kms.
}
Figure \ref{fig:aux-Functions} shows that the deviations from this maximum are
no more than a factor of \about\ 2 when \tV\ is varied in either direction by
two orders of magnitude. The dependence of velocity on luminosity is only
$L^{1/4}$, and since $L_4$ is typically \about\ 0.3--20 it introduces a
similarly small variation. Finally, even the dependence on grain properties is
weak, as is evident from the expression for $c_3$ (eq.\ \ref{c3-4}). The only
grain parameter that enters linearly is the condensation temperature, and its
dispersion is expected to be small.

%%%%%%%%%%%%%%%%%%% Auxilary functions %%%%%%%%%%%%%%%%%%%%%%%%%%%%%
\begin{figure}
 \centering \leavevmode \psfig{file=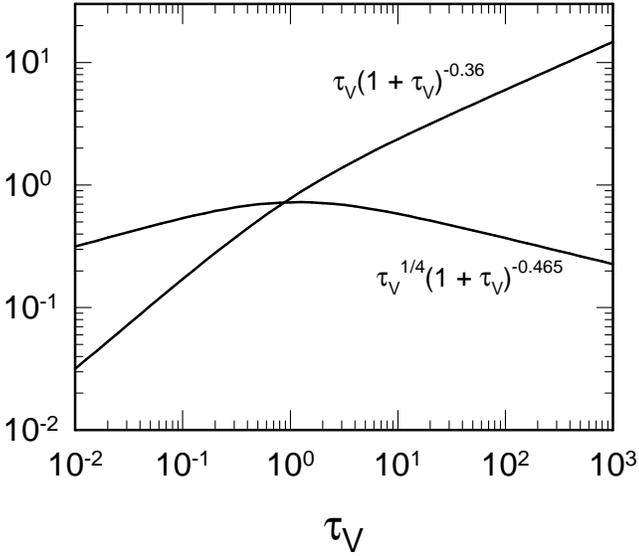,width=\figsize,clip=}
\caption{The \tV-variation of the correlations \ref{C2} and \ref{C4}.
} \label{fig:aux-Functions}
\end{figure}
%%%%%%%%%%%%%%%%%%%%%%%%%%%%%%%%%%%%%%%%%%%%%%%%%%%%%%%%%%%%%%%%%%%%%%%%%%%%

Since only two of the $K$-functions are independent, the universal fits for
$K_3$ and $K_4$ can be used to produce grain-independent approximations also
for $K_1$ and $K_2$. Figure \ref{fig:K-Functions} shows that the single
function $K_3^3/K_4 = (1 + \tV)^{0.105}$ provides an adequate fit for $K_1$ of
both silicate and carbon dust, always within 30\% of the detailed
results\footnote{Higher accuracy, when desired, can be obtained by replacing
the common index 0.105 with 0.05 for amorphous carbon and 0.15 for silicate
grains.}. This expression gives
\eq{\label{C3}
 \Msix = c_1\,\left(L_4\tV\right)^{3/4}\left(1 + \tV\right)^{0.105}. \\
}
Since \tV\ = 1.3 gives the largest possible velocity at a given $L$, the
corresponding mass loss rate is
\eq{\label{Mmax}
    \Msix(v\sub{max}) = 1.33\,c_1\,L_4^{3/4}.
}
A common practice in analysis of observations is to deduce the optical depth
from spectral data and derive \Mdot\ from \tV\ (e.g., Jura 1991). This
procedure is predicated on the assumption of a linear relationship between
\Mdot\ and \tV, and our result shows that this assumption is not quite right.
At a fixed luminosity, $\Mdot$ is proportional to $\tau_V^{3/4}$ in the
optically thin regime and to $\tau_V^{0.86}$ at large \tV. The expectation of a
linear relationship between \Mdot\ and \tV\ is not met because of the dust
drift, since $\tV/\Mdot \propto \int(\zeta/y^2v)dy$. The relationship would be
linear if $\zeta$ were 1 everywhere, but figure \ref{fig:radial} shows that
this is never the case. The deviations from unity are large at small \tV, where
the drift is most prominent, decreasing as \tV\ increases.

Finally, figure \ref{fig:K-Functions} shows that the universal profile
$K_3^4/K_4 = (1 + \tV)^{-0.36}$ again describes reasonably the actual $K_2$ of
both grains, so that
\eq{\label{C4}
 \Mdot\vf = {L\over c}\left(\Qast/\QV\right)\,\tV\left(1 + \tV\right)^{-0.36}
}
The relation $\Mdot\vf \le L/c$ has often been used as a physical bound on
radiatively driven winds, even though the mistake in this application when $\tV
> 1$ has been pointed out repeatedly. In IE95 we show that the proper form of
momentum conservation is $\Mdot\vf = \tF L/c$ where \tF\ is the flux-averaged
optical depth, so now we have found the explicit expression
\eq{\label{tauF}
        \tF = \left(\Qast/\QV\right)\,\tV(1 + \tV)^{-0.36}.
}
The \tV-variation of this function is shown in figure \ref{fig:aux-Functions}.
It increases linearly at small \tV, reaches unity at \tV\ = 1.6 and switches to
$\tau^{0.64}\sub V$ thereafter. The ratio \Qast/\QV\ is 0.1 for silicate and
0.25 for carbon (see Table 1).

For every pair among \Mdot, $L$ and \vf, equations (\ref{C1}), (\ref{C2}) and
(\ref{C3}) list the correlation in term of optical depth.  When detailed IR
data are not available and \tV\ cannot be determined, it can be bypassed by
correlating the pairs directly against each other.  Since only two of the three
relations are independent, there is only one such combination. Equation
\ref{C3} is the most suitable for eliminating \tV\ because $K_1$ varies the
least among the four $K$-functions. In fact, the crude approximation $K_1
\simeq 1$ introduces an error of less then 50\% for $\tV \la 10$.  With this
approximation, $\tV \propto \Mdot^{4/3}/L$.  Inserting this result in equation
\ref{C1} yields
\eq{\label{universal}
    v_1^3 = A\,\Msix\left(1 + B\,{\Mdot_{-6}^{4/3}\over L_4}\right)^{-1.5}
}
where $A = c_4$ and $B = c_1^{-4/3}$. This universal correlation summarizes our
solution for all dusty winds away from the boundaries of phase space. The error
in this result, introduced by the approximation $K_1 = 1$, is less than 50\%
when $\tV \le 10$. When the observational accuracy warrants higher precision,
corrections can be readily derived.

This completes the similarity solution of the dusty wind problem.  Our results
amplify the conclusion of IE95, taking it a step further: The solution is fully
characterized by optical depth. The relations among global parameters \Mdot,
$L$ and \vf\ involve universal similarity functions of \tV, independent of
chemical composition. The grain properties enter only in the proportionality
constants of the similarity relations. We derived the similarity functions from
solutions for carbon and silicate grains, whose absorption efficiencies are
widely different. Since dust spectral features have a negligible effect on
overall reddening corrections, these functions should describe all interstellar
grains with reasonable properties. It is gratifying that in spite of its great
complexity, the dusty wind problem can afford such a simple, explicit solution.

\subsection{Young's Correlation}
\label{Young}

Young (1995) conducted a survey of nearby Mira variables with low mass-loss
rates.  He finds a clear, strong correlation between outflow velocity and
mass-loss rate, but independent of luminosity. The correlation can be
parametrized as $\Mdot \propto v_\infty^\alpha$, with $\alpha = 3.35$.
Subsequent observations by Knapp et al (1998) corroborate Young's results and
find $\alpha$ = 2, although the scatter in their data is consistent with values
as large as 3. Remarkably, even though the wind is driven by radiation
pressure, its velocity is independent of luminosity.

From equation \ref{C1} (or \ref{universal}), at small optical depths our
solution gives $\Mdot \propto \vfff$ independent of luminosity, explaining the
observational findings. The implied $\alpha = 3$ is consistent with all
observational results within their errors. This result reflects the central
role of drift at small mass loss rates. When the drift dominates, $\zeta \simeq
w^{1/2} \propto (\Mdot v/L)^{1/2}$. Neglecting gravity and reverting to
physical variables, the equation of motion (\ref{force a}) becomes
\eq{
    {dv^2\over dr} \propto {(\Mdot L v)^{1/2}\over r^2}
}
where the proportionality constant contains the grain properties. The solution
gives $v_\infty^{3/2} \propto (\Mdot L)^{1/2}/r_1$. And since $r_1$ is
proportional to $L^{1/2}$ (eq.\ \ref{r1}), the dependence on luminosity cancels
out. If not for the particular $L$-dependencies of $r_1$ and the drift, this
would not have happened.

The observed correlation, in particular its luminosity independence, directly
reflects the specific dependence of the drag force on \Mdot, $L$ and $v$. Other
forms for the force would produce different correlations. For example, when the
drift is neglected, the combination $(\Mdot L v)^{1/2}$ is replaced by $L$,
leading to the GS result $\vf = \vp \propto (L/r_1)^{1/2}$ (eq.\ \ref{vp}).
Together with $r_1 \propto L^{1/2}$ this yields $v_\infty^4 \propto L$, as
first noted by Jura (1984).  This prediction of a correlation between velocity
and luminosity independent of mass-loss rate is in strong conflict with the
observations. One could formally eliminate the luminosity with the aid of the
momentum flux conservation $\Mdot \vf = \tF L /c$ (see eq.\ \ref{C4} and
subsequent discussion) to re-write this result as $\Mdot \propto \tF \vfff$.
While this bears superficial resemblance to Young's correlation, the
observational result emerges only if the variation of \tF\ with $L$ and \Mdot\
is ignored. By contrast, at $\tV < 1$ equations \ref{C3} and \ref{tauF}
together give $\tF \propto \tV \propto \Mdot^{4/3}/L$.

Young's correlation is a direct reflection of the basic physics ingredients
that went into the model. It demonstrates the importance of drift in dusty
winds and provides strong support for its underlying theory.

\section{PHYSICAL DOMAIN}
\label{sec:boundaries}

A global property not considered thus far is the stellar mass.  This quantity
does not enter into the definition of the parameter $P$ and thus cannot be
determined in general. The mass only affects $\Gamma$ (see equation
\ref{Gamma}), a quantity that can vary by orders of magnitude without a
discernible effect on any observed property. However, $\Gamma$ does play an
important role in determining the parameter range corresponding to actual
winds.

The equation of motion (\ref{equation}) has many mathematical solutions but not
all of them are physically relevant.  By example, consider the analytic
solution for optically thin winds (equation \ref{CompleteSolution}), whose
logarithmic term becomes singular when $\delta = \dmax$  (see equation
\ref{dmax}). The singularity is avoided whenever the numerator and denominator
are finite and have the same sign.  Both cases are mathematically acceptable
but the negative sign is physically meaningless.  Such solutions violate the
liftoff condition (equation \ref{liftoff}) and the only transition from one set
of solutions to the other is through the singularity.

At the outset, $P$, $\Gamma$ and $\theta$ must be positive.  Further, all
winds must obey the liftoff condition, which involves only two out of the
three input parameters. This condition ensures liftoff under all circumstances
but it does not automatically guaranty a meaningful outcome. The formal
solution of equation \ref{equation} gives
\eq{
  \wff = \theta^2 + P^2\left(\int_1^{\infty}\phi\zeta{dy\over y^2} -
      {1\over\Gamma}\right).
}
Obviously, in physical solutions the final velocity must exceed the initial
velocity and we further require $\vf/\vT \ga 3$ (see \S\ref{sec:thin}).
Because $\phi \le 1$ and $\zeta < 1$, $\int_1^{\infty}(\phi\zeta/y^2)dy < 1$
and
\eq{\label{bound2}
    P^2 > \theta^2{\Gamma\over\Gamma - 1}\left({\vff\over\vTT} - 1\right).
}
The three parameters of every dusty wind whose velocity increases by the ratio
\vf/\vT\ must also obey this constraint.

%%%%%%%%%%%%%%%%%%% Q-Profiles %%%%%%%%%%%%%%%%%%%%%%%%%%%%%
\begin{figure}
 \centering \leavevmode \psfig{file=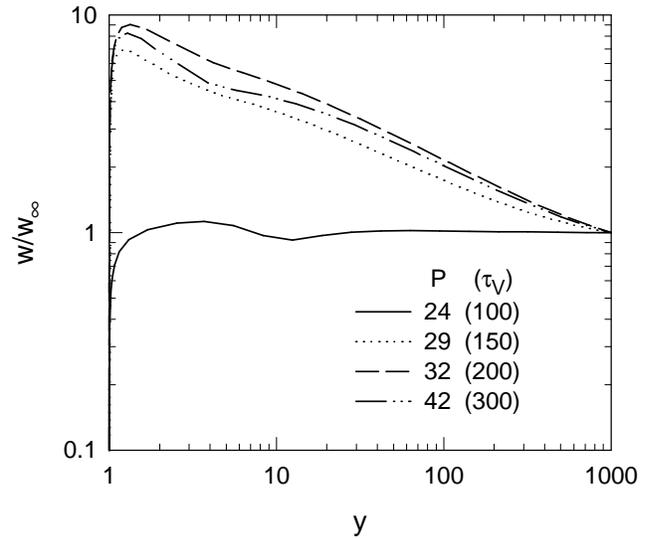,width=\figsize,clip=}
\caption{Gravitational quenching because of reddening. Models are for silicate
dust with the listed values of $P$ (the corresponding \tV\ is in parenthesis).
In each case $\theta$ and $\Gamma$ are adjusted so that $\wf\ = 10\,\theta$ and
the gravitational pull reaches 99\% of its possible maximum; the detailed
procedure is described in Appendix \ref{numerics}. } \label{fig:quenched}
\end{figure}
%%%%%%%%%%%%%%%%%%%%%%%%%%%%%%%%%%%%%%%%%%%%%%%%%%%%%%%%%%%%%%%%%%%%%%%%%%%%

\subsection{Gravitational Quenching}
\label{Quenching}

The wind acceleration is positive as long as $\Gamma\phi\zeta > 1$ (eq.\
\ref{equation}). Negative acceleration, leading to wind quenching, can be
caused by increasing either the gravitational pull (smaller $\Gamma$) or the
reddening (smaller $\phi$). The quenching process depends on all three input
parameters and takes different forms in the optically thick and thin regimes.
In optically thin winds $\phi$ = 1 and since $\zeta$ increases monotonically
with $w$ (cf figure \ref{fig:radial}), the liftoff condition ensures positive
acceleration everywhere. Consider the analytic solution (eq.\
\ref{CompleteSolution}) when $\delta$ increases while $P$ and $\theta$ remain
fixed. When $\beta = \delta/\dmax$ approaches unity the acceleration becomes
negligible, though it remains positive, producing the behavior seen in the
plots for $\beta$ = 0.999 and 0.99999 in figure \ref{fig:thin}. This quenching
effect is caused by the logarithmic term. Near $y = 1$ this term dominates and
the velocity increases very slowly. At a distance $y$ roughly proportional to
$P^{-2}\theta^{3/2}|\ln(1 - \beta)|$, the terms independent of $\delta$ take
over and the acceleration picks up. From that point on the solution resembles
non-quenched winds, albeit at a lower acceleration because of its late start.
As $\beta$ gets closer to unity, the dominance of the logarithmic term is
slowly extended further out until the whole wind is stalled.

Reddening introduces an entirely different quenching mode, affecting winds that
obey the liftoff condition when $P$ is increased beyond a certain point. The
liftoff condition ensures that the initial acceleration is always positive,
irrespective of optical depth. However, subsequent reddening can reduce $\phi$
substantially (figure \ref{fig:radial}) and since the acceleration is
proportional to $\phi\zeta - 1/\Gamma$, the wind may be prevented from reaching
a significant terminal velocity. Furthermore, since $\phi\zeta < \phi$ the
acceleration becomes negative whenever $\phi < 1/\Gamma$ and the velocity can
even decline after its initial rise. Figure \ref{fig:quenched} shows examples
of such winds whose velocity profile differs greatly from the monotonic rise
that typifies all solutions not too close to the boundary of phase space. In
each case \wf\ = 10$\,\theta$. But when $P \ga 25$, the velocity reaches
intermediate values substantially higher than \wf, almost as much as ten times
higher, before declining to its final value. This behavior has intriguing
observational implications, but it is not clear whether the limited phase space
for such solutions affords a meaningful number of cases.

\subsection{Phase Space Boundaries}

Every point in the 3-dimensional $P$-$\Gamma$-$\theta$ space that results in a
physical solution can be labeled by \wf\ of that solution. The equation $\wf(P,
\Gamma, \theta) = C\theta$, where $C \ge 3$ is some prescribed number, defines
a 2-dimensional surface in the solution space.  This surface is the locus of
solutions whose velocity increases by factor $C$, i.e., every wind on this
surface has $\vf/\vT = C$. The volume enclosed by this surface corresponds to
solutions with $\vf/\vT \ge C$, solutions outside this volume have $\vf/\vT <
C$.

Figure \ref{fig:boundary} shows the projections onto the $P$-$\Gamma$ plane of
two such surfaces with representative values of \vf/\vT. Each curve encloses
the allowed region for winds whose \vf/\vT\ is at least as large as the
boundary mark. These boundaries are best understood by considering the wind
problem at a fixed $\Gamma$. Moving from the bottom of the figure along a line
of constant $\Gamma$ and increasing $P$, the curve for a given \vf/\vT\ is
first intersected at its lower branch. This intersection defines a minimal $P$
for physical solutions, which can be estimated by inserting $\theta$ from the
liftoff condition into equation \ref{bound2}; a more accurate expression for
this $P\sub{min}$ follows from the analytic solution for optically thin winds.
The region below the lower branch is forbidden for that particular \vf/\vT.
Increasing $P$ further crosses into the allowed region, leading to winds whose
\vf/\vT\ exceeds the boundary mark. Eventually, quenching by reddening sets a
maximum to the value of $P$ that still allows \vf/\vT\ as large as the boundary
mark. This \Pmax\ corresponds to the second intersection with the curve, and
the region above the upper branch again is forbidden. The two branches of each
\vf/\vT\ boundary are defined by the two gravitational quenching modes and
reflect the central role of $P$ in controlling both the dynamics and radiative
transfer; this parameter sets the scales for both the acceleration (eq.\
\ref{equation}) and optical depth (eq.\ \ref{eta_tV}). The physical domain for
$P$ is between the two branches, $P\sub{min} < P < P\sub{max}$. As $\Gamma$
decreases, the gravitational force increases and the two branches approach each
other: $P\sub{min}$ becomes larger (it takes stronger acceleration to overcome
gravity) and $P\sub{max}$ becomes smaller (it takes less reddening to quench
the wind).  The meeting point of the two branches defines an absolute minimum
for $\Gamma$, determined exclusively by the grain properties. Below this
minimum, liftoff requires such a high $P$ that reddening quenches the wind
immediately. Note that this minimum is about 3--5 times larger than the
Eddington limit $\Gamma$ = 1, showing that the latter is not a sufficient
condition for radiatively driven winds when the dust drift is taken into
account (see below).

The plots in figure \ref{fig:boundary} outline the proper phase space
boundaries for radiatively driven winds, replacing the erroneous bound
$\Mdot\vf \le L/c$. If the velocity of a wind increases by a certain ratio, the
parameters $P$ and $\Gamma$ of that wind necessarily lie inside the region
bounded by the curve plotted for that ratio. The parameter $\theta$ is not
shown in this projection but must conform to the bounds expressed by the
liftoff condition and equation \ref{bound2}.

%%%%%%%%%%%%%%%%%%% Boundary %%%%%%%%%%%%%%%%%%%%%%%%%%%%%
\begin{figure}
 \centering \leavevmode \psfig{file=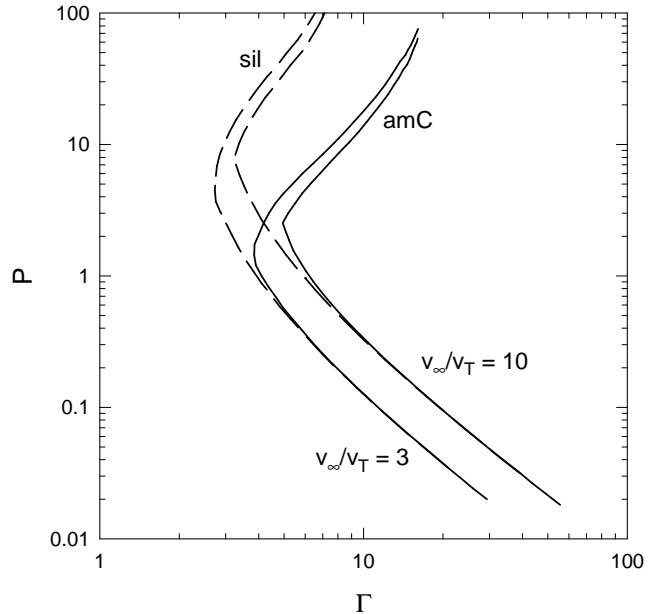,width=\figsize,clip=}
\caption{Phase space boundary: The parameters $P$ and $\Gamma$ of winds whose
velocities increase by the indicated ratios \vf/\vT\ must lie to the right of
the correspondingly marked boundaries. The boundaries for amorphous carbon are
marked by amC, for silicate grains by sil. \label{fig:boundary}}
\end{figure}
%%%%%%%%%%%%%%%%%%%%%%%%%%%%%%%%%%%%%%%%%%%%%%%%%%%%%%%%%%%%%%%%%%%%%%%%%%%%

\subsubsection{The Lower Branch; Minimal \Mdot}

This branch reflects the constraints imposed by equation \ref{bound2} and the
liftoff condition. Equation \ref{bound2} implies that $\wf < P$ in all winds,
namely $v < \vp$.  As we discussed in \S\ref{Theory}, observations usually give
$v \ll \vp$. While this inequality poses a problem for the simple GS solution,
the proper inclusion of drift and reddening shows that it actually provides
important support for radiation pressure as the driving mechanism in dusty
winds.

From the liftoff condition (\ref{liftoff}), all winds must obey $\Gamma > 1$,
the standard Eddington limit.  Figure \ref{fig:boundary} shows that in dusty
winds this bound is superseded by the combined effects of drift and reddening,
which together impose a more stringent lower limit on $\Gamma$. The Eddington
limit would be meaningful if the dust and gas were position coupled, instead
the relevant regime in AGB winds is momentum coupling (see Gilman 1972).
Indeed, $\Gamma > 1$ implies
\eq{
             {L_4 \over M_0}\,\Qast\ss > 2.18\x\E{-2}
}
and this condition is obeyed by a large margin with typical parameters of
observed winds.  Although the Eddington limit does not impose a meaningful
bound it provides another consistency check on the basic premises of the model.

The liftoff condition also implies that
\eq{
    \theta > \case1/4
        \left[\left(\Gamma + 3\over\Gamma - 1\right)^{1/2} - 1\right]^2
        \simeq \cases{(\Gamma - 1)^{-1} & when $\Gamma \to 1$  \cr
                                                                        \cr
                      \Gamma^{-2}       & when $\Gamma \gg 1$   }
}
Since $\Gamma$ cannot approach unity (figure \ref{fig:boundary}), the
meaningful constraint is $\theta > \Gamma^{-2}$. From its definition in
equation \ref{w-P-theta}, $\theta = \Mdot/\Mc$ where
\eq{
    \Mc = {\Qast L\over\vT c}
        = 7.06\x\E{-5}\,{\Qast L_4\over T^{1/2}\sub{k3}}\ \Mo\ \rm yr^{-1}
}
is a characteristic mass loss rate defined by the free parameters.  Therefore,
the liftoff condition implies a lower bound on the mass loss rate
\eq{\label{MinM0}
    \Mdot > {\Mc\over\Gamma^2}\,.
}
When this constraint is violated, the grains are ejected without dragging the
gas because the density is too low for efficient momentum transfer from the
dust to the gas.

The practical application of this result requires care. This constraint
directly reflects our handling of the wind origin, the least understood part of
the problem. Our liftoff condition is predicated on the assumption of positive
acceleration right from the start. However, the initial acceleration in fact is
negative in the two-fluid formulation that neglects gas pressure; starting such
a calculation with $v = \vd$ produces a negative $dv/dr$, as is evident from
both the NE (their equation 36) and HTT (eq.\ 7) studies. The wind is still
lifted in these calculations because the dust accelerates, reversing the
initial gas infall. But the negative derivative in turn reflects the
shortcomings of these studies since the gas is in hydrostatic equilibrium in
the absence of dust. This deficiency is removed in the study by Kwok (1975)
which includes also gas pressure. At the sonic point Kwok finds ${\cal
F\sub{rad}/\cal F\sub{grav}} < 1$ (his equation 15), in direct conflict with
our liftoff assumption; the relation between the two forces is reversed only
after further grain growth. But Kwok's result, too, is unrealistic because it
neglects the effect of radiation pressure on the gas molecules, likely to be
instrumental in initiating the outflow (e.g.\ Elitzur, Brown \& Johnson 1989).

An exact calculation of the minimal \Mdot\ is beyond the scope of our study. It
would require proper inclusion of physical processes, such as grain growth,
whose understanding is still rudimentary at best. It is also instructive to
recall that even the formulation of drag is only accurate to within ``\ldots
factors of order 2 or 3, which depend on accommodation coefficients and
geometric factors" (Salpeter 1974). Fortunately, these uncertainties involve
only the very initial stage of the outflow and the severity of their impact is
greatly reduced once the flow settles into the highly supersonic stage that is
the main thrust of our study. Indeed, the minimal dependence on $\theta$ of our
final results demonstrates that for the most part, the initial conditions are
quite irrelevant in the supersonic regime. While $\Gamma^2\theta > 1$ remains a
strict requirement of our model, in reality this liftoff constraint could be
less restrictive without an appreciable effect on most of our conclusions.

We propose a phenomenological substitute for the mathematical liftoff
constraint
\eq{\label{fudge}
    \Gamma^2\theta \ga f,
}
where $f\ (< 1)$ is an unknown factor. This only affects results that directly
involve the liftoff relation --- the location of the lower branch of the
phase-space boundary (figure \ref{fig:boundary}) and the estimate of the lower
bound on \Mdot\ (equation \ref{MinM0}). We estimate the phenomenological factor
$f$ by noting that it is mostly relevant at small $P$. In that region our
solution gives $w_\infty^3 = \case9/{16}P^4$ (equation \ref{wf2}), which can be
rewritten as
\eq{
  \Gamma^2\theta = \case{16}/9\,\left(\vT\over\vg\right)^{4}
                        \left(\vf\over\vT\right)^{3}.
}
With nominal values for \vT\ and \vg, the requirement $\vf/\vT \ga 3$ and
consistency of the last two relations set $f$ \about\ 0.1. As a result,
\eq{
 \Mdot \ga f{\Mc\over\Gamma^2} \simeq
 3\x\E{-9}{M_0^2\over\Qast\ssq L_4T^{1/2}\sub{k3}} \ \Mo\ \rm yr^{-1}
}
replaces equation \ref{MinM0} as our phenomenological estimate of the minimal
\Mdot. It may be noted that HTT proposed a similar relation.

\subsubsection{The Upper Branch; Maximal \Mdot}

The upper branch of the phase space boundary is defined by the quenching effect
of reddening. Since it involves the wind outer regions where the supersonic
flow is well established, it does not suffer from the liftoff uncertainties
that afflict the lower branch. The upper branch defines a maximum \Mdot\ when
the other parameters are fixed, discussed in paper II.

\section{GRAIN MIXTURES}
\label{Mixtures}

We now discuss the general case of a mixture of grains that can have different
sizes and chemical compositions. For the most part, the results of the
single-type case hold. A new feature is the variation of the mixture
composition because of different drift velocities.

The $i$-component of the mixture is defined by its grain size $a_i$ and
efficiency coefficients $Q_{i,\lambda}$; the latter may reflect differences in
both size and chemical composition.  As before, the details of production
mechanism are ignored. We define the wind origin $y = 1$ as the point where the
last grain type added to the mix enables the conditions for momentum coupling,
initiating the supersonic gas outflow. We assume no further change in grain
properties and denote the density of species $i$ at that point $n_i(1)$ and its
fractional abundance $\xi = n_i(1)/\nd(1)$.

\subsection{Scaling Formalism}

In analogy with the single-type case, the reddening profile of the i-th species
is
\eq{
 \phi_i(y) = {1 \over Q_{\ast,i}}\int \Qi{F_\lambda(y)\over F(y)}\,d\lambda,
}
where
\[
 Q_{\ast,i} = {\pi\over\sigma T_\ast^4}\int\Qi B_\lambda(T_\ast)d\lambda\,.
\]
Since grain-grain collisions are negligible, different types of grains drift
through the gas in response to the radiation pressure independent of each
other. The velocity of species $i$ is $v_i$ and the ratio $n_i/\nH$ varies in
proportion to $\zeta_i(y) = v/v_i$. Introduce
\eq{\label{eq:qi}
    q_i = {Q_{\ast,i}\over\Qast}\,,   \qquad
    \hbox{where}\
    \Qast = \sum_i\xi Q_{\ast,i}\,.
}
With the aid of the mixture average \Qast\ we define \vm\ as before (equation
\ref{vm}), and the drift function of the $i$-th species is
\eq{\label{eq:zeta_i}
  \zeta_i(y) = {\theta + \sqrt{wq_i\phi_i}\over
                \theta + q_i\phi_i + \sqrt{wq_i\phi_i}}\,,
}
where $w$ and $\theta$ are the same as in the single-type case (eq.\
\ref{w-P-theta}). The radiation pressure force per unit volume is
\eq{
  {\cal F}\sub{rad}
       = {L\Qast\over4\upi r^2 c}\sum_i n_i(y)\upi a_i^2 q_i\phi_i(y).
}
We generalize the definition of cross-section area per gas particle (equation
\ref{sg}) via
\eq{
    f_i\sg = \upi a_i^2\,{\xi\over\zeta_i(1)}\,{\nd(1)\over\nH(1)}
    \qquad \hbox{and} \    \sum f_i = 1,
}
so that $f_i$ is the fractional contribution of species $i$ to \sg. At every
point in the shell $n_i\upi a_i^2 = \nH\sg f_i\zeta_i$ and the equation of
motion is
\eq{
   {dw^2\over dy} = {P^2\over y^2}\left[\sum_i f_i q_i\phi_i(y)\zeta_i(y)
    - {1\over\Gamma}\right]
}
where $P$ and $\Gamma$ are defined as before. The equation of motion retains
its single-type form, the only modification is the weighted sum in the
radiation pressure term.  The dust properties require now a larger amount of
input but the other parameters remain the same --- $P$, $\theta$ and $\Gamma$.

The radiative transfer equation requires as input the overall optical depth
\tV\ and dust density profile $\eta$. Introduce the normalized density profile
of the i-th species
\eq{
  \eta_i(y) = {\zeta_i(y)\over y^2 w(y)}
       \left(\int_1^\infty\!\!\! {\zeta_i\over y^2 w}\, dy\right)^{\!\!\!-1}
}
and the coefficients
\eq{
    z_i = Q\sub{Vi} f_i \int_1^\infty\!\!\! {\zeta_i\over y^2 w}\, dy
}
where $Q\sub{Vi}$ is the absorption efficiency at visual of the $i$-th
species. Then
\eq{
    \eta = {\sum_i z_i\eta_i\over\sum_i z_i}, \qquad
    \tV  = \case1/2P^2\sum_i z_i
}
and the radiative transfer problem again does not require any additional input
parameters.  The most general dusty wind problem is fully prescribed by the
grain properties and the three free parameters $P$, $\theta$ and $\Gamma$.

\subsection{The Wind Velocity}

The wind velocity profile remains similar to the single-type case. Consider
first optically thin winds, where $\phi_i = 1$. An analytic solution does not
exist now but the problem reverts to the single-grain case in two limits. When
$w$ is sufficiently small that  $w < q_i$ for every $i$, drift dominates all
grain types and the $k = \case2/3$ profile of eq.\ \ref{profile1} is recovered.
In the opposite limit of $w > q_i$ for every $i$, drift is negligible for all
grains and eq.\ \ref{profile1} again applies with $k = \case1/2$. In reality,
the grain types are distributed between these two extremes.  Different types
move from the drift-dominated to the negligible-drift regimes as $w$ increases,
and the gas velocity profile evolves slowly from one shape to the other. Since
the two profiles are quite similar, the overall behavior does not differ
significantly from the single-grain case.  And since the only effect of
reddening is to decrease $k$ slightly further, the qualitative similarity
remains for optically thick winds.

\subsection{Radial Variation of Grain Abundances}

Grains of different types have different velocities, creating a radial
variation of the fractional abundances. The final abundance of species $i$ is
$\xi E_i/E$ where
\eq{\label{eq:Ei}
   E_i = {\zeta_i(\infty)\over\zeta_i(1)}
}
and $E = \sum\xi E_i$. Since $\zeta$ is monotonically increasing, $E_i > 1$ for
every grain type and $E > 1$.  The outflow enhances the overall dust abundance,
the final value of \nd/\nH\ always exceeds its initial value.  The fractional
abundance is enhanced for grains with $E_i >E$ and suppressed for those with
$E_i < E$.

For simplicity, we discuss the behavior of this differential enhancement only
in the optically thin limit ($\phi_i = 1$), where $E_i$ has the approximate
behavior
\eq{
   E_i \simeq \cases
 {1                                   &$\phantom{111111}    q_i < \theta$ \cr
 \DS\left(q_i\over\theta\right)^{1/2} &$\phantom{11}\theta < q_i < \wf$   \cr
 \DS\left(\wf\over\theta\right)^{1/2} &$\wf                < q_i $         }
}
This result is easy to understand. At a position where the gas velocity is
$w$, grains with $q_i < w$ maintain a constant ratio $v_i/v$ and have $\zeta_i
= 1$. Those with $q_i > w$ are falling behind and have $\zeta = (w/q_i)^{1/2}$.
The expression for $E_i$ reflects the two limiting cases of grains that are
either moving with the gas ($q_i < \theta$) or lagging behind ($q_i > \wf$)
throughout the entire outflow, and the intermediate case of grains that are
falling behind until the wind has accelerated to the point that $w = q_i$.

The differential enhancement is controlled by $q_i$, the ratio of \Qast\ for
the particular grain type and for the whole mixture.  The dependence of this
quantity on grain size can be derived in a simple model that ignores spectral
features, approximating the efficiency coefficient with
\eq{
    Q(\lambda,a) \simeq \cases{1                   &$\lambda < 2\upi a$
                                                                    \cr\cr
       \DS\left(2\upi a\over\lambda\right)^\beta   &$\lambda > 2\upi a$}
}
This crude approximation, with $\beta$ \about\ 1--2, is adequate for most
purposes. Approximating the spectral averaging in the definition of \Qast\ with
its value at the peak wavelength of the Planck distribution produces
\eq{
    Q_{\ast}(a) \simeq \cases{1                   &$a > a_\ast$
        \cr\cr
       \DS\left(a\over a_\ast\right)^\beta        &$a < a_\ast$
        }
}
where $a_\ast$ = 0.64\,\mic\,(1000 K/$T_\ast$). It is interesting to note that
$T_\ast$ = 2500 K gives $a_\ast$ = 0.25 \mic, same as the upper bound of the
MRN grain size distribution (Mathis, Rumpl \& Nordsieck 1977).

Consider an initial power-low grain size distribution of the form
\eq{
  \nd(a, y = 1) \propto a^{-p} \qquad \hbox{for} \quad \am \le a \le a_\ast
}
defined by the free parameters $p$ and \am\ (the MRN distribution corresponds
to the special case with $p = 3.5$ and \am\ = 0.005 \mic). With this general
form,
\eq{
        q(a) = \left(a\over a_1\right)^\beta,     \qquad \hbox{where}\
        a_1 = \am\left(p - 1\over p - \beta - 1\right)^{1/\beta}.
}
Only the lower bound \am\ enters here, $q(a)$ is independent of the upper bound
$a_\ast$. The result for the abundance enhancement factor is
\eq{
    E(a) \simeq \cases{
     1         &$\phantom{a_1\theta^{1/\beta} < } a < a_1\theta^{1/\beta}$  \cr
     \DS \left(a\over a_1\theta^{1/\beta}\right)^{\beta/2}
                           &$a_1\theta^{1/\beta} < a < a_1\wf^{1/\beta}$  \cr
     \DS \left(\wf\over \theta\right)^{1/2}
                           &$a_1\wf^{1/\beta} < a$                     }
}
The drift enhances preferentially the upper end of the size distribution,
grains smaller than a certain size do not drift at all and their abundance is
never enhanced. {\em A power-law distribution that starts with index $p$
emerges as a power-law distribution with the reduced index $p - \beta/2$}. To
produce the MRN index of 3.5, the size distribution must start with a higher
index of \about\ 4--4.5.

A complete theory for the grain distribution cannot be constructed without
including the details of the production process. Our results show that some
properties of the MRN distribution are natural features in the subsequent
effect the outflow has on the fractional abundances. It is possible that the
MRN distribution could emerge as the only self-consistent solution of the
complete theory.

\section{SUMMARY AND DISCUSSION}
\label{summary}

Given grain properties, we have shown here that the supersonic phase of dusty
winds is specified in terms of three independent dimensionless variables. A
bank of solutions derived in the 3-dimensional space of these variables will
contain the dimensionless density and velocity profiles of any possible wind
around an evolved star. Matching the complete observations of any particular
system requires at most scaling by an overall velocity scale. In addition, the
solution of the radiative transfer problem is also contained in the outcome.
We find that the solutions are generally controlled by a single parameter and
can be described in terms of self-similarity functions that are independent of
grain properties. Therefore, the results presented here contain the solution
for the dynamics problem for all grains. The corresponding spectral energy
distributions were presented elsewhere for amorphous carbon and silicate dust
(IE95, IE97). Thanks to scaling, extending that database to any other grain
composition is as simple as a single run of DUSTY in which \tV\ is varied over
its entire feasible range.

Our formulation starts once the dust properties are established. These
properties, as well as the value of \Mdot, are set during an earlier phase of
the outflow. This initial phase is still poorly understood and our analysis has
nothing to say about it. However, most observations involve the subsequent
supersonic phase, which is the one addressed here. We associate the starting
radius $r_1$ with prompt dust formation, but this does not enter directly into
the solution because $r_1$ drops out of our formulation. Extended, rather than
prompt, dust formation can be accommodated without any change if $r_1$ is
identified with the endpoint of grain growth. The association of $r_1$ with
dust condensation enters only indirectly---since the region $r < r_1$ is devoid
of dust, the diffuse radiation vanishes at $r_1$ (IE97).  Even this indirect
effect is inconsequential in optically thin systems, where the dynamics impact
of the diffuse radiation is negligible altogether. Therefore, {\em our analytic
solution eq.\ \ref{CompleteSolution} is valid in all optically thin winds
irrespective of the nature of dust formation}; if grain growth is extended over
a distance comparable with the dynamics length scale, the solution simply
becomes applicable from the point where that growth stops. The only direct
reference to the association of dust condensation with $r_1$ occurs when this
radius is expressed in terms of other physical quantities (eq.\ \ref{r1}). If
the explicit dependence on $r_1$ is instead left intact, any arbitrary model of
dust condensation can be incorporated through the behavior it gives for $r_1$.
Young's correlation (sec.\ \ref{Young}) implies that $r_1 \propto L^{1/2}$,
consistent with prompt dust formation.

One of the grain parameters employed here is the cross-section area per
hydrogen nucleus \sg. This quantity is directly related to the ratio of
dust-to-gas mass loss rates
\eq{
    \rdg = {\Mdot\sub{d}\over\Mdot}
         = {4\rho\sub{s}\over3\mp}\,a\sg
}
where $\rho\sub{s}$ is the density of the dust solid material. This ratio
(which requires additionally $\rho\sub{s}$) was used as an input in the
calculations of HTT, who denoted it $\delta$ and pointed out its significance
to the wind solution. Our analysis shows that \sg\ actually does not enter into
the scaling formulation of the problem; indeed, we have solved here the dusty
wind problem for two different types of grains without ever specifying the
value of \sg\ for either of them.  The only dust properties required are the
efficiency coefficients $Q_\lambda$ and sublimation temperature \Tc.  Even the
grain size is not strictly needed since it only enters indirectly as one
factor, albeit a major one, in determining $Q_\lambda$.  The dust abundance is
an extraneous parameter controlling only the correspondence between the model
results and physical quantities, just like the velocity scale which is
irrelevant for the solution itself.

Our analysis highlights the central roles that drift and reddening play in the
velocity structure of dusty winds.  Independent of grain properties, the
significance of drift is diminished when $P > \case{16}/9$, a universal result
reflecting the mathematical structure of the equation of motion. Reddening, on
the other hand, becomes important at $\tV > 1$ and the corresponding $P$ is
determined by \QV\ (equation \ref{P1}). If \QV\ were, say, \E{-2} reddening
would start playing a role only at $P > 100$, long after the drift ceased to be
a factor. Since astronomical dust has \QV\ of order unity, reddening starts
controlling the force just as the drift dominance ends.

Finally, why are typical velocities of AGB winds \about\ 10--20 \kms? Our
results provide explanations for both the magnitude and the narrow range of
\vf. While the driving force is radiation pressure, drift and reddening play a
fundamental role. Rapid drift at small \tV\ decreases the dust abundance from
its initial value, reducing the captured fraction of radiative energy. And at
large optical depths this fraction is similarly reduced because the radiation
spectral shape is shifted toward longer wavelengths where the absorption
efficiency is small. The two effects combine to produce the very narrow range
of velocities displayed by equation \ref{C2} --- the range of variation with
either $L$ or \tV\ is less than factor of 3, the dependence on dust properties
is similarly weak. Given that the luminosity scale is \about\ \E4\ \Lo, the
velocity magnitude is determined by the dust parameters. Single grain
properties cannot deviate much from the values listed in Table 1, the only
parameter that is completely free is the dust abundance. Since the velocities
are typically \about\ 10 \kms, the dust abundance must be such that \ss\ is of
order unity; were that abundance 100 times higher, typical velocities would be
\about\ 100 \kms\ instead. This implies that the dust is produced with an
abundance comparable to that found in the general interstellar medium and
resolves the conflicts pointed out by Castor (1981).

\section*{Acknowledgments}

We thank Haggai Elitzur for instrumental advice on the analytic solution of
optically thin winds.  The partial support of NASA and NSF is gratefully
acknowledged.

\appendix
\section{Glossary}
\label{app:Glossary}

\def\e#1{\smallskip\noindent\parbox[t]{0.25\columnwidth}{#1}}
\def\d#1{\parbox[t]{0.7\columnwidth}{#1}}

\e{$a$} \d{dust grain radius}

\e{$E_i$} \d{abundance enhancement of the $i$-th component of a mixture owing
to differential drift velocities (eq.\ \ref{eq:Ei})}

\e{$K_1 \dots K_4$} \d{similarity functions summarizing the solution observable
implications (equations \ref{K1}, \ref{K2} and \ref{K3-4})}

\e{$L, L_4$} \d{luminosity and its magnitude in \E4~\Lo}

\e{$M, M_0$} \d{stellar mass and its magnitude in \Mo}

\e{$\Mdot, \Msix$} \d{mass loss rate and its magnitude in \E{-6} \Mo\
yr$^{-1}$}

\e{$\nd, \nH$} \d{number densities of dust grains and hydrogen nuclei,
respectively}

\e{$P = \vp/\vm$} \d{main scaling parameter of the dimensionless equation of
motion; sets the ratio of radiation pressure to the drift effect}

\e{$\QV$} \d{the dust absorption efficiency at visual}

\e{$Q_\ast$} \d{Planck average at the stellar temperature of the efficiency
coefficient for radiation pressure (eq.\ \ref{q*})}

\e{$q_i$} \d{fractional contribution of the $i$-th component of a mixture to
\Qast\ (eq.\ \ref{eq:qi})}

\e{$r_1, r_{1,14}$} \d{the wind inner radius and its magnitude in \E{14} cm}

\e{$T_\ast$} \d{stellar temperature}

\e{$\Tc, \Tct$} \d{the dust temperature at $r_1$ and its magnitude in 1,000 K}

\e{$\Tk, \Tkt$} \d{the kinetic temperature at $r_1$ and its magnitude in 1,000
K}

\e{$v$, \vd, \vrel} \d{velocities of the gas, dust and their difference,
respectively}

\e{$v\sub{p}$, \vm, \vg} \d{velocity scales characterizing the radiation
pressure (eq.\ \ref{vp}), drift (\ref{vm}) and gravity (\ref{vg}),
respectively}

\e{$v\sub{T}$} \d{the wind initial velocity (identified with the isothermal
sound speed at $r_1$; eq.\ \ref{vT})}

\e{$v_\infty, v_1$} \d{the wind final velocity  and its magnitude in \kms}

\e{$w = v/\vm$} \d{dimensionless velocity}

\e{$w_\infty = v_\infty/\vm$} \d{the dimensionless final velocity}

\e{$y = r/r_1$} \d{dimensionless radial distance}

\e{$\Gamma$} \d{ratio of radiation pressure to gravity (eq.\ \ref{Gamma})}

\e{$\delta = 1/(\Gamma - 1)$} \d{auxiliary quantity for the analytic solution
of optically thin winds}

\e{$\zeta = v/\vd$} \d{dimensionless drift profile (eq.\ \ref{zeta})}

\e{$\zeta_i$} \d{dimensionless drift profile of the $i$-th component of a gas
mixture (eq.\ \ref{eq:zeta_i})}

\e{$\eta$} \d{dimensionless dust density profile (eq.\ \ref{eta_def})}

\e{$\theta = \vT/\vm$} \d{the dimensionless initial velocity}

\e{$\Theta$} \d{the transformation function between $P$ and \tV\ (eq.\
\ref{Theta0})}

\e{$\sigma\sub{g}, \ss$} \d{dust cross-section area per particle at
condensation and its magnitude in \E{-22} cm$^2$ (eq.\ \ref{sg})}

\e{$\tau\sub{V}$} \d{overall optical depth at visual}

\e{$\phi$} \d{reddening profile (eq.\ \ref{phi})}

\e{$\Phi$} \d{velocity-weighted harmonic average of $\phi$ (eq.\ \ref{Phi})}

\e{$\bar \phi$} \d{density-weighted average of $\phi$ (eq.\ \ref{Psi})}

\e{$\Psi, \Psi_0$} \d{a dimensionless function determined by radiative transfer
and its value in optically thin winds (eqs.\ \ref{r1} and \ref{Psi0})}

\section{Drift Velocity}
\label{app:drift}

The drag force on a grain moving at velocity \vrel\ through gas with density
$\rho$ is
\eq{
  F\sub{drag} =  \pi a^2\rho\,\x\cases{v^2\sub{rel}  & when \vrel\ $>$ \vT  \cr
                           \vrel\vT  & when \vrel\ $<$ \vT   }
}
for the subsonic (\vrel\ $<$ \vT) and supersonic (\vrel\ $>$ \vT) regimes
(e.g., Kwok 1975). The drift reaches steady state when $F\sub{drag} =
F\sub{pr}$, the radiation pressure force on the grain. Introduce
\eq{
    \vDD = {F\sub{pr}\over\pi a^2 \rho}
         = {1\over c\rho}\int\!\Q F_\lambda\,d\lambda,
}
then the steady-state drift velocity is
\eq{
    \vrel = \cases{\vD             & when \vD\ $>$ \vT  \cr
                                                        \cr
                   {\vDD\over\vT}  & when \vD\ $<$ \vT   }
}
Both limits can be combined in the simple expression
\eq{\label{vdrift}
    \vrel = {\vDD\over\vD + \vT}.
}
Kwok proposed instead the more complex expression
\eq{
 \vrel = \left\{\case1/2\left[\left(4v^4\sub D + v^4\sub T\right)^{1/2}
        - v^2\sub T\right]\right\}^{1/2}.
}
He derived this result by combining first into a single expression the two
limiting forms of the drag force, instead of \vrel\ itself, and then solved the
resulting quadratic equation for \vrel.  Kwok's expression and ours are based
on the same ingredients and give identical results in the limits of both \vD\
$>$ \vT\ and \vD\ $<$ \vT.

\section{Analytic Approximations}
\label{analytic}

The quantities of interest obtained from the solution of the equation of motion
are the profile shapes of the velocity and density variation, the final
velocity \wf\ and the optical depth \tV.  Introduce
\eq{
    u = {w\over\wf},                        \qquad
    \epsilon = {\theta\over\wf},               \qquad
    \beta   = {\delta\over\dmax}\,.
}
The profile shape is conveniently expressed in terms of $u$, which varies from
$\epsilon$ to 1, with $\epsilon\ (\le \case1/3)$ and $\beta\ (< 1)$ as small
parameters. Furthermore, we note that $\theta$, too, is a small parameter in
all practical cases and use $\dmax = \theta + \sqrt{\theta} \simeq
\sqrt{\theta}$ as well as $\delta - \theta \simeq \beta\sqrt{\theta}$.

\subsection{Negligible Reddening}

When $\phi$ = 1, the complete solution (eq. \ref{CompleteSolution}) gives the
following equation for \wf:
\eqarray{
    P^2 &= &\wff\left(1 - \epsilon^2\right)
                \left(1 + \beta\epsilon^{1/2}w^{1/2}_\infty\right)    \non
 &\phantom{11}+& \case4/3w^{3/2}_\infty
            \left(1 + \beta\epsilon^{1/2}w^{1/2}_\infty\right)^2\times\non
 &&\quad \bigg[1 + \case3/2\beta\epsilon^{1/2} - 3\beta^2\epsilon
         - \epsilon^{3/2}\left(1 + \case3/2\beta + 3\beta^2\right)   \non
 &&\qquad+\
     3\beta^3\epsilon^{3/2}\ln{1 - \beta\epsilon^{1/2}
                          \over\epsilon^{1/2}(1 - \beta)} \bigg].
}
This expression is an expansion in the problem's small parameters. Consider
first the limit $\beta = 0$, which gives
\eq{\label{wf1}
    P^2 = \case4/3w^{3/2}_\infty\left(1 - \epsilon^{3/2}\right)
        + \wff\left(1 - \epsilon^2\right).
}
At small values of \wf\ the first term on the right dominates, at large values
the second. Keeping only the dominant term produces
%$w^3_\infty = \case9/{16}P^4\,(1 - \epsilon^{3/2})^{-2}$
$\wf = (\case9/{16}P^4)^{1/3}\,(1 - \epsilon^{3/2})^{-2/3}$  in the first
regime, $\wf = P(1 - \epsilon^2)^{-1/2}$ in the second. Neglecting the
$\epsilon$ corrections, the two terms and the approximate solutions they yield
are equal to each other at $\wf = P = \case{16}/9$. Since small \wf\ ensures $w
< 1$ for the entire wind, $\zeta = w^{1/2}$ in that case while $\zeta = 1$ for
most of the wind when $\wf \gg 1$ (see equation \ref{equation}).  Therefore, $P
= \case{16}/9$ is the transition between drift dominance and negligible drift,
and
\eq{\label{wf2}
 \wf = \cases{
       \left(\case9/{16}P^4\right)^{1/3} & for $P < 16/9$ (drift dominated)\cr
                \cr
       P & for $P > 16/9$ (negligible drift)}
}
These two limit expressions for \wf\ can be combined in the single form listed
in equation \ref{wf0}.

Finite $\beta$ corrections are generally small when $\epsilon\ \le \case1/3$
and $\beta \le \case1/2$. The deviations from unity of the terms inside the
large square brackets are less then 30\% in that domain. The only significant
corrections can come from the terms in $\beta\epsilon^{1/2}w^{1/2}_\infty$.
Since $\wf \le P$, $\beta\epsilon^{1/2}w^{1/2}_\infty < 0.3P^{1/2}$ for all
$\epsilon \le \case1/3$ and $\beta \le \case1/2$. Therefore, this term is
always negligible in the drift dominated regime. But at large $P$ it can become
significant, at $P = 10$ it already amounts to a 90\% correction.  When this
term dominates, the solution becomes $w^5_\infty = P^4/\beta^2\epsilon$.
However, reddening effects become significant before this limit is reached,
therefore we are justified in maintaining the $\beta$ = 0 approximation; figure
\ref{fig:thin} shows that this faithfully presents the entire $\epsilon \le
\case1/3$ and $\beta \le \case1/2$ domain.

We proceed now to the velocity profile. With $\beta$ = 0, equation
\ref{CompleteSolution} gives
\eq{
    {u^2 - \epsilon^2 + \case4/3w^{-1/2}_\infty(u^{3/2} - \epsilon^{3/2})\over
       1 - \epsilon^2 + \case4/3w^{-1/2}_\infty(1 - \epsilon^{3/2})}
   = 1 - {1\over y}\, .
}
Similar to equation \ref{wf1}, this equation changes its behavior at $\wf =
\case{16}/9$; at smaller \wf\ the terms proportional to
$\case4/3w^{-1/2}_\infty$ dominate, at larger \wf\ the other terms. The two
limits give
\eq{\label{u}
    u = \left(1 - {1 - \epsilon^{1/k}\over y}\right)^k,     \qquad
    k = \cases{2/3 & for $P < 16/9$ \cr
                   &            \cr
               1/2 & for $P > 16/9$}
}
From this we can immediately determine the dust density normalized profile
$\eta\ (\propto \zeta/y^2u)$, required for the radiative transfer equation.
The ratio $\zeta/u$ is proportional to $1/u^{1/2}$ in the drift-dominated
regime and to $1/u$ when the drift is negligible. With the particular value for
$k$ in each case, $\zeta/u$ is the same as $1/u$ with the power index $-k$
replaced by $1 - k$, leading to
\eq{\label{eta}
    \eta = {A\over y^2}
        \left({y\over y - 1 + \epsilon^{1/k}}\right)^{\!\!\!1 - k}, \qquad
        A = k{1 - \epsilon^{1/k}\over 1 - \epsilon}\,.
}
The velocity and density profiles are independent of $P$, which is relevant
only for the choice of the solution regime. The only parameter that enters
explicitly is $\epsilon$, and even this dependence is confined mostly to the
wind origin where $u(1) = \epsilon$ and $\eta(1) = A/\epsilon^{(1 - k)/k}$.
However, the $\epsilon$-dependence, which must be maintained to avoid a
singularity for $\eta$ at $y$ = 1, rapidly disappears once $y > 1 + \epsilon$.

\subsection{Reddening Corrections}

Reddening corrections are conveniently expressed in terms of the quantity
$\Phi$. Since $\phi$ decreases away from the wind origin, the averaging is
dominated by its upper end and
\eq{
    \Phi \simeq {\phi(y \to \infty)\over1 - \epsilon}\,.
}
For simplicity, only $\delta = 0$ is considered. The equivalent of equation
\ref{wf1} is then
\eq{
    P^2 = \case4/3w_{\infty}^{3/2}\int_{\epsilon}^1{du\over\sqrt{\phi}} +
          \wff\int_{\epsilon}^1{du\over\phi}.
}
Approximating the integrand in the first term by its largest value, our
approximate solution for \wf\ becomes
\eq{\label{wf-P}
   P^2  = {4w_{\infty}^{3/2}\over3\sqrt{\Phi}} + {\wff\over\Phi}\,.
}
In analogy with equation \ref{wf1} this yields
\eq{
 \wf = \cases{
    \left(\case9/{16}P^4\Phi\right)^{1/3} & for $P < \case{16}/9\Phi^{1/2}$\cr
               \cr
       P\Phi^{1/2} & for $P > \case{16}/9\Phi^{1/2}$ }
}

\subsection{Optical Depth}

The relation between $P$ and \tV\ is obtained by combining equations
\ref{tV_work} and \ref{wf-P}. When $\delta = 0$,
\eq{\label{P-tauV}
    {P^2\over\Phi} =
    {4\over3}\left(\tV\over\QV\right)^{3/2} + \left(\tV\over\QV\right)^2.
}
From this we can find the value of $P$ that yields \tV\ = 1. With the
approximation $\Phi \simeq 1$ at this point we get
\eq{
    P = {1\over\QV}\left(1 + \case4/3Q^{1/2}\sub{V}\right)^{1/2}.
}
Finally, the $\delta$-dependent correction term in equation \ref{tV_work} can
be estimated in the optically thin case. With the leading-order velocity
profile from equation \ref{u}, the integration result is
\eq{
 \int_1^\infty{dy\over y^2u} = {1\over1 - k}\,
    {1 - \epsilon^{(1 - k)/k}\over 1 - \epsilon^{1/k}}.
}

\section{Numerical Procedures}
\label{numerics}

The velocity and density profiles are very steep near the wind origin (cf eq.\
\ref{eta}). As a result, the solution of the equation of motion \ref{equation}
can become a difficult numerical problem that requires a prohibitive number of
radial grid points, especially when optical depths are large. To bypass these
problems, we find the velocity profile instead from the integral equation
expressed by the formal solution
\eq{\label{eq1}
    w^2 = \theta^2 + P^2\left[z(y) -
                {1\over\Gamma}\left(1 - {1\over y}\right)\right],
}
where
\eq{
    z = \int_1^y\phi\zeta{dy\over y^2}\,.
}
Given $\phi$, this equation is solved by iterations that start with the
analytic solution for $w$ derived in the previous section.

The code DUSTY (Ivezi\' c et al, 1999) starts by solving the radiative
transfer equation with an initial density profile $\eta$ taken from equation
\ref{eta}.  The resulting reddening profile $\phi$ is used in equation
\ref{eq1} to find $w$. A new reddening profile is calculated with the
corresponding $\eta$ (see equation \ref{eta_tV}), and the process is repeated
until $\phi$, $\eta$ and $w$ are self-consistent. Convergence is rapid and the
number of radial grid points is modest --- less than 30 points are adequate
for optical depths up to 100. The power of the method stems from the fact that
the steepness of the velocity and density profiles is built in right from the
start.

We find it advantageous to replace $P$, $\Gamma$ and $\theta$ with three other
independent input parameters. First, because of the central role of radiative
transfer we specify the optical depth \tV\ and determine $P$ from
\eq{
    P^2 = {2\over N}{\tV\over\QV}\,,
    \qquad \hbox{where}\
    N = \int_1^\infty\!\!\!{\zeta \over y^2w}\ dy
}
(see equation \ref{eta_tV}). Next, consider the quantity
\eq{
    \gm = \max{\DS 1 - {1\over y }\over z(y)}\,,
}
the maximum of the prescribed profile anywhere in the shell. The wind stalls
when $\gm = \Gamma$, therefore physical solutions have $\gm = f\Gamma$ with $f
< 1$. We choose $f$ as an input parameter instead of $\Gamma$ with the
replacement
\eq{
    {1\over\Gamma} = {f\over\gm}\,.
}
The parameter $f$ varies from 0 for negligible gravity to 1 for
gravitationally quenched winds. Finally, instead of $\theta$ we specify as an
input parameter $\epsilon = \theta/\wf$. If $Y$ denotes the shell outer radius,
$\theta$ is derived from the formal solution at that point, which gives
\eq{
    \theta^2 = {\epsilon^2\over1 - \epsilon^2}\,P^2
        \left[z(Y) - {1\over\Gamma}\left(1 - {1\over Y}\right)\right].
}
The advantage of \tV, $\epsilon$ and $f$ over $P$, $\theta$ and $\Gamma$ as
input parameters is the more direct contact they have with physical
quantities. In particular, the choice $\epsilon \le \case1/3$ and $f < 1$
guarantees a priori a physically meaningful solution.

\label{lastpage}
\end{document}